\documentclass[12pt,final]{iopart}
\usepackage{hyperref}
\hypersetup{
    colorlinks=true,   
    linkcolor=blue,    
    citecolor=blue,    
    filecolor=magenta, 
    urlcolor=cyan      
}
\usepackage{calrsfs}
\usepackage{iopams}

\DeclareMathAlphabet{\pazocal}{OMS}{zplm}{m}{n}

\usepackage{bm}
\usepackage{bbm}
\usepackage{revsymb}
\usepackage[perpage]{footmisc}

\usepackage [english]{babel}
\usepackage [utf8]{inputenc}
\usepackage [T1]{fontenc}
\usepackage{graphicx}
\usepackage{setstack}
\usepackage{siunitx}
\usepackage{booktabs}

\usepackage{xcolor}

\def\dvec#1{ \overset{\text{\tiny$\bm\leftrightarrow$}}{#1} }
\newcommand{\Tensor}[1]{ \dvec{\mathbf{#1}} }

\newcommand\Id{\mathbbm{1}}  

\newcommand \ket [1]{ | #1 \rangle }
\newcommand \bra [1]{ \langle #1 | }
\newcommand \ketbra [2]{ \ket{#1} \! \bra{#2} }

\renewcommand \Re { \mathop{\mathrm{Re}} }
\renewcommand \Im { \mathop{\mathrm{Im}} }

\newcommand \Trace { \mathop{\mathrm{Tr}} }

\makeatletter
\protected\def \ConfAvg {\@ifstar\@ConfAvg@resize@auto\@ConfAvg@check@option}
\def \@ConfAvg@check@option {\@ifnextchar[\@ConfAvg@resize@fixed\@ConfAvg@resize@no}
\def \@ConfAvg@resize@no #1{
  \langle
    #1
  \rangle
  \sb {\mathrm{conf}}
}
\def \@ConfAvg@resize@auto #1{
  \left\langle
    #1
  \right\rangle
  \sb {\mathrm{conf}}
}
\def \@ConfAvg@resize@fixed [#1]#2{
  \csname\expandafter\@gobble\string#1l\endcsname\langle
    #2
  \csname\expandafter\@gobble\string#1r\endcsname\rangle
  \sb {\mathrm{conf}}
}

\providecommand*\clap[1]{\hb@xt@\z@{\hss#1\hss}}
\providecommand*\mathclap{
  \expandafter \mathpalette \expandafter \MTmathclapNn
}
\def\MTmathclapNn #1#2{{}\clap{$\m@th#1{#2}$}}

\makeatother

\newcommand \uvec[1]{ \hat{\mathbf{#1}} }

\newcommand{\rmx}{x}
\newcommand{\rmy}{y}
\newcommand{\rmz}{z}

\newcommand{\rmg}{\mathrm{g}}
\newcommand{\rmL}{\mathrm{L}}
\newcommand{\ProjE}{P_{\rme}}
\newcommand{\ProjG}{P_{\rmg}}
\newcommand{\JE}{J_{\rme}}
\newcommand{\JG}{J_{\rmg}}

\newcommand{\la}{\langle}
\newcommand{\ra}{\rangle}
\newcommand{\Du}{\textbf{D}^{\dagger}}
\newcommand{\Dd}{\textbf{D}}
\newcommand{\vk}{\textbf{k}}

\newcommand{\FLI}{\mathcal{F}}  

\newcommand{\Mcyc}{M_{\mathrm{cyc}}}
\newcommand{\Tcyc}{T_{\mathrm{cyc}}}

\newcommand{\dif}[1]{\mathinner{\rmd #1}}

\newcommand{\Liou}{ \mathcal{L} }

\newcommand{\LiouG}{ \Liou_\gamma }
\newcommand{\LiouL}{ \Liou_{\mathrm{L}} }
\newcommand{\LiouInt}{ \Liou_{\mathrm{int}} }

\newcommand{\Eps}{ \boldsymbol{\varepsilon} }
\newcommand{\EpsL}{ \Eps_{\mathrm{L}} }

\begin{document}
\date{\today}

\title{Theory of multiple quantum coherence signals in dilute thermal gases}

\author{Benedikt Ames$^1$\footnote{Present address: Department of Physics and Materials Science,
		Campus Limpertsberg, Universit\'e du Luxembourg 
		162 A, avenue de la Fa\"iencerie, 
		L-1511 Luxembourg}, Edoardo G.~Carnio$^{1,2}$, Vyacheslav N.~Shatokhin$^{1,2}$, and Andreas Buchleitner$^{1}$} 

\address{$^1$ Physikalisches Institut, Albert-Ludwigs-Universit\"at Freiburg, Hermann-Herder-Str.\ 3, D-79104 Freiburg, Germany}
\address{$^2$ EUCOR Centre for Quantum Science and Quantum Computing, Albert-Ludwigs-Universit\"at Freiburg, Hermann-Herder-Str.\ 3, D-79104 Freiburg, Germany}

\begin{abstract}
Manifestations of dipole-dipole interactions in dilute thermal gases are
difficult to sense because of strong inhomogeneous broadening. Recent
experiments reported signatures of such interactions in fluorescence detection-based measurements of multiple quantum coherence (MQC) signals,
with many characteristic features hitherto unexplained. 
We develop an original open quantum
systems theory of MQC 
in dilute thermal gases, which allows us to
resolve this conundrum.
Our theory accounts for the vector character of the atomic dipoles
as well as 
for driving laser pulses of arbitrary strength, 
includes the
far-field coupling between the dipoles,
which prevails in dilute ensembles, and effectively incorporates atomic motion 
via a disorder average.
We show that collective decay processes -- which were ignored in previous
treatments employing the electrostatic form of dipolar
interactions -- play a key role in the emergence of MQC signals.
\end{abstract}


\section{Introduction}
Dipole-dipole interactions enable transport and
collective phenomena like dipole blockade in
Rydberg gases \cite{PhysRevLett.87.037901},
coherent backscattering of light in cold atoms \cite{labeyrie99},
resonant energy transfer in photosynthetic complexes
\cite{Brixner:2005jb,shatokhin2018_njp}, interatomic Coulombic decay
\cite{Cederbaum1997,Hemmerich:2018wd}, sub- \cite{PhysRevLett.108.123602} and
superradiance \cite{PhysRevLett.117.073003}, and multipartite entanglement
\cite{platzer10,sauer2012}.
A typical manifestation of dipolar interactions are collective 
displacements of the
energy  levels of a many-body system \cite{Friedberg1973101}.
However, in thermal ensembles the observation of these 
is hindered by
motionally induced Doppler shifts.
As a result, a
mean-field description 
is often sufficient \cite{PhysRevLett.112.113603},
and the properties of the system can be derived from single-body responses.

Such picture is however not truly complete,
since signatures of light-induced dipolar interactions indeed can be
observed in so-called \emph{multiple quantum coherence} (MQC)
signals \cite{Dai_2012} sensing coherent superpositions of the ground and
collective excited states of several particles.
While originally MQC signals were measured \cite{Dai_2012} via ultrafast
two-dimensional electronic spectroscopy \cite{Jonas2003}, which allows one to
probe directly the nonlinear susceptibilities responsible for MQC signals, more
recently these signals were extracted from fluorescence emitted by dilute
thermal atomic gases \cite{lukas_bruder15}. 

The interpretation of fluorescence-detected MQC
signals sparked a debate.
On the one hand, it was pointed out~\cite{Mukamel_2016} that
fluorescence as radiated by dilute atomic samples can usually be well understood
by considering a system of independent emitters, which corresponds to a
mean-field description.
On the other hand, the theoretical follow-up \cite{Li_2017} to the experiment
\cite{lukas_bruder15} predicted vanishing double-quantum coherence in the
absence of interactions, inferring that the latter are essential to establish
collective behaviour.
A subsequent experiment~\cite{Bruder_2019} further reported significant
quantitative deviations from the results based on the model of~\cite{Li_2017}.
Eventually, it was suggested~\cite{Bruder_2019} that 
the double-quantum
coherence signals could only be understood by going beyond a two-body model---into the realm of many ($N>2$) interacting particles.

The aim of our present article is to settle this debate by elucidating the role of
dipole-dipole interactions in MQC signals derived from
fluorescence-based measurements. 
These signals are emitted when atoms are pumped and probed by a series of
femtosecond laser pulses, and, until now, their interpretation
\cite{Dai_2012,lukas_bruder15,Li_2017} relies on two orthodox premises  of
multi-dimensional nonlinear spectroscopy~\cite{mukamel_book,PhysRevA.49.146}:
(a) the laser-atom interactions are sufficiently weak and can be accounted for
perturbatively, using double-sided Feynman diagrams, and (b) the dipole-dipole
coupling scales as $\sim R^{-3}$ with the interatomic distance $R$, and
corresponds to the familiar \emph{electrostatic}~\cite{Jackson}, or \emph{near-field}
interaction associated with \emph{virtual} photons~\cite{Andrews04}.

The theory of MQC signals we present here abandons both these premises. 
Since fluorescence is an \emph{incoherent} process, resulting from spontaneous decay
of excited emitters (be them independent or interacting),
its dynamics
can be described with the
formalism of open quantum systems \cite{breuer_book}.
More specifically, we adapt a quantum-optical master equation approach
previously used to faithfully model coherent backscattering (CBS) of light by
cold atoms \cite{shatokhin2005,shatokhin2006,ketterer2014,binninger19}.
Unlike the diagrammatic approach, our method (i) treats the laser-matter
interaction \emph{non-perturbatively}, and (ii) considers the \emph{exact} form of
dipolar coupling, which, beyond the electrostatic part, includes the exchange of photons
via its \emph{radiative}, or \emph{far-field} component, scaling as $\sim (k_0R)^{-1}$, with $k_0=2\pi/\lambda_0$ the wave number ($\lambda_0$ the wave length) 
resonant with the 
individual atomic scatterers' electronic transition frequency $\omega_0$.
This latter extension (ii) in particular ensures the inclusion of irreversible,
\emph{collective} radiative decay processes, which are ignored by the electrostatic 
modelling \cite{Jackson} of the dipolar coupling.

\begin{figure}
  \center
  \includegraphics[width=0.5\columnwidth]{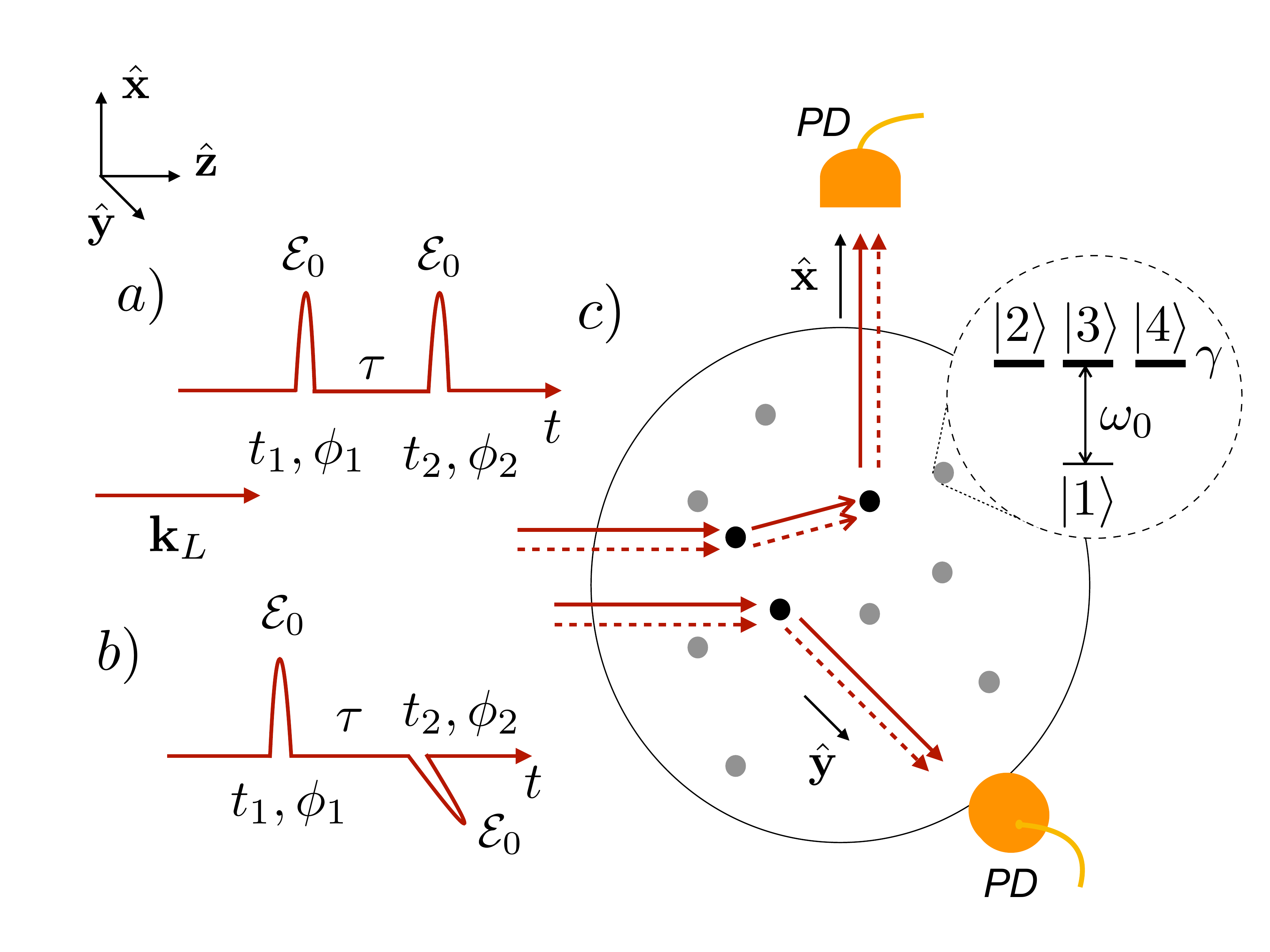}
  \caption{%
    Setup for the observation of multiple quantum coherence (MQC) signals.
    Two phase-tagged (by phases $\phi_1$ and $\phi_2$), femtosecond, pump and probe laser pulses with variable
    time delay $\tau =t_2-t_1$ (also see fig.~\ref{fig:timescales}) propagate along the $\rmz$ axis defined by the wave vector
    ${\mathbf k}_{\rmL}$.
    The pulses have (a) parallel linear polarizations along the $\rmx$-axis; (b)
    orthogonal linear polarizations along the $\rmx$- and $\rmy$-axes, respectively.
    (c) Disordered sample of identical atomic scatterers with electronic dipole
    transition $\JG = 0 \rightarrow \JE = 1$ at frequency $\omega_0$, and 
    spontaneous decay rate $\gamma$ of each excited state sublevel $\ket{2}$, $\ket{3}$, $\ket{4}$.
    Solid arrows represent incident and scattered field amplitudes, 
    dashed ones 
    the associated complex conjugates. 
    Sequences of red arrows   
    and black dots illustrate single and double
    scattering processes that contribute to the fluorescence signals 
    collected by the
    photodetectors (PD) in the observation directions $\uvec{y}$ and $\uvec{x}$,
    respectively.
  }
  \label{fig:setup}
\end{figure}

The radiative part of the dipole-dipole interaction prevails in dilute gases,
where the mean interatomic separation is much larger than the resonant optical
wave length.
Since $(k_0R)^{-1}\ll 1$ in this regime, we can treat this interaction
perturbatively.
Then the average fluorescence intensity can be represented by a series expansion in powers
of $(k_0R)^{-1}$, whose subsequent terms correspond (see fig.~\ref{fig:setup}) to single, double, etc.\ %
scattering contributions.
We consequently interpret MQC components distilled from the total
fluorescence signal as arising from the multiple scattering of real photons between
distant atoms.

Below, we recall the basics of the phase-modulated spectroscopy that was
employed to measure MQC signals in dilute thermal gases
\cite{lukas_bruder15,Bruder_2019,Bruder_phd}.
Thereafter, we present our formalism and derive 1QC (single quantum coherence)
and 2QC (double quantum coherence) spectra that are in excellent qualitative
and reasonable quantitative agreement  with experiment
\cite{Bruder_2019}.
Ultimately, we establish the crucial role of radiative dipole-dipole interactions in
the emergence of 2QC signals, which \emph{are impossible to observe for independent
atoms} -- in agreement with \cite{Li_2017}, and we show that a system of two dipole interacting atoms is sufficient for
modelling 1QC and 2QC signals -- as a consequence of double scattering of real photons -- in a dilute thermal gas.
We especially highlight 
collective decay processes as an essential part of
the dipolar interactions, without which qualitative agreement with the
experiment cannot be achieved. 

We now delve into a thorough presentation of the theory; the reader who is already familiar with the topic may take a shortcut to \sref{sec:spectra}, where we present our results.


\section{Phase-modulated spectroscopy}
\label{sec:spectroscopy}

\begin{figure}
	\center
	\includegraphics{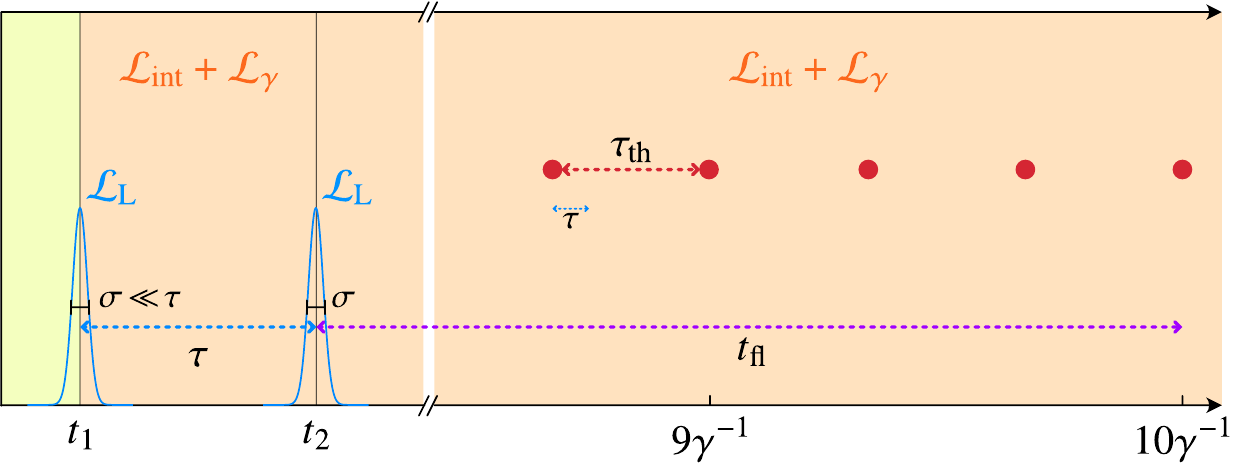}
	\caption{Schematic illustration of the different time scales and of the associated Liouvillians, for a single experimental cycle of duration $\Tcyc$. Two ultrashort Gaussian laser pulses with the same width $\sigma$ arrive at times $t_1$ and $t_2$, separated by a delay $\tau\gg \sigma$. After the second pulse, fluorescence is monitored during the interval $t_\mathrm{fl}\sim 10\gamma^{-1}\gg \tau$, until all atoms decay into the ground state. The fluorescence detection time $t_\mathrm{fl}$ is much longer than the thermal decoherence time $\tau_{\rm th}>\tau$, beyond which the atomic motion cannot be ignored. During the ultrashort laser-atom interaction, the dynamics of the atomic operators $Q(t)$ in \eref{meq} is governed by the Liouvillian $\LiouL$. Instead, $\LiouInt$ and $\LiouG$ generate the dynamics 
	during 
	the intervals 
	$\tau$ and $t_\mathrm{fl}$.}
	\label{fig:timescales}
\end{figure}

In \fref{fig:setup} we sketch the spectroscopy experiment
used to extract the MQC signals.
A thermal ensemble of alkali atoms is excited near-resonantly
\cite{lukas_bruder15,Bruder_2019} by two subsequent laser pulses
injected at times $t_1$ and $t_2>t_1$, respectively
\cite{Tekavec2006}.
The fluorescence subsequently emitted by the sample is 
collected by a photodetector in
a direction orthogonal to the exciting field's wave vector ${\bf k}_{\rm L}$. This excitation and detection procedure is
repeated for $\Mcyc \gg 1$ cycles of duration $\Tcyc$, 
with the two pulses 
garnished with distinct 
phase tags $\phi_1$ and $\phi_2$, respectively, which are linearly ramped over subsequent cycles, with distinct rates $w_1\ne w_2$. These slowly evolving 
phase tags 
will turn out to be instrumental 
to distill
the 
MQC signals from the total fluorescence. 
$\Tcyc$ is chosen to significantly 
exceed
the atomic
radiative lifetime
of a few tens of nanoseconds~\cite{steckRubidium87Line2015}, such that the atoms decay to
the ground state~\cite{lukas_bruder15,Bruder_2019} within a single cycle, and each of the $\Mcyc$~cycles
can be considered in isolation (see fig.~\ref{fig:timescales}).

To resolve the details of the excitation scheme, 
we focus
on
the $m$th cycle ($0 \leq m < \Mcyc$),
which begins at $\tau_m = m \Tcyc$. 
For positive delays 
$t \in [0, \Tcyc ]$ with respect to $\tau_m$,
and in the frame co-rotating with the laser carrier frequency
$\omega_{\rmL}$,
the injected field has the time-dependent amplitude
\begin{equation}
  {\mathbf E}_{\rmL}(\mathbf{r},t)
  =
  \sum_{j=1}^2
    \EpsL^{(j)}
    {\cal E}_{\rmL}(t - t_j)
    \rme^{ \rmi(
      {\mathbf k}_{\rmL} \cdot \mathbf{r}
      + \omega_{\rmL}t_j 
     + \phi_j
    )}
  \, ,
\label{laser}
\end{equation}
with the same Gaussian envelope $
  {\cal E}_{\rmL}(t')
  =
  {\cal E}_0 \exp(-t'^2 / 2 \sigma^2)
$ 
of both incoming pulses, 
amplitude ${\cal E}_0$ and duration $\sigma$. 
Each 
phase tag $\phi_j = w_j (\tau_m+t)$
can be approximated by $\phi_j\approx w_j \tau_m$
for 
modulation frequencies $w_j\ll \omega_L$ \cite{Tekavec2006} \footnote{%
	The 
	phase tags~$\phi_j$
	vary from cycle to cycle, but
	we drop the cycle index~$m$ 
	to 
	keep notation slim.
},
and is imprinted on the respective pulse
by transmission
through
an
acousto-optical modulator oscillating at 
frequency
$w_j$. 
Both
pulses are collinear, with wave vector $\mathbf{k}_{\rmL}$, but may have different polarizations
$\EpsL^{(j)}$: in the following we 
consider the cases $\EpsL^{(1)} = \EpsL^{(2)} = \uvec{x}$, as well as 
 $\EpsL^{(1)} = \uvec{x}$ and $\EpsL^{(2)} = \uvec{y}$. 
In a typical experiment, 
$\sigma \sim \SI{50}{\femto\second} \ll 
\tau = t_2 - t_1 \sim \SI{10}{\pico\second} \ll
\Tcyc \sim \SI{300}{\micro\second}$.

As a consequence of the phase tagging of the driving field, the transient fluorescence
intensity $I_{\uvec{k}}(\tau, t_{\rm fl}, \phi_1,\phi_2)$ in the direction $\uvec{k} \perp {\bf k}_{\rm L}$
exhibits a modulation -- in its dependence on $\tau_m$ through $\phi_1$ and $\phi_2$ -- at 
the 
frequency difference
$w_{21}=w_2-w_1$, and at higher harmonics thereof, with the $\kappa$th harmonic identifying 
the $\kappa$th
quantum coherence ($\kappa$QC) \cite{lukas_bruder15}.
So does the 
time-integrated (over the $m\rm th$ cycle)
fluorescence signal as collected 
in the far field 
over times $t_2+t_{\rm fl}$, with 
$0\leq t_{\rm fl} \leq \Tcyc-t_2\gg \gamma^{-1}$, 
$\gamma$ the natural line width of the atomic excited states, 
faithfully approximated by
\begin{equation}
  \bar{I}_{\uvec{k}}(\tau, \phi_1,\phi_2)
  =
  \int_{0}^{\infty} \dif{t_{\rm fl}}
    I_{\uvec{k}}(\tau,t_{\rm fl},\phi_1,\phi_2)
  \, .
  \label{intensity}
\end{equation}
When incident on a photodetector, the 
integrated signal generates a photocurrent 
proportional to $\bar{I}_{\uvec{k}}(\tau,\phi_1,\phi_2)$, which (independently)
depends on the pulse delay $\tau$, and -- via the phase tags $\phi_1,\phi_2$ (see above) -- on the cycling time $\tau_m$. 
Frequency analysis of the photocurrent with respect to 
$\tau_m$, to distill the targeted order $\kappa$ of MQC, and with respect to $\tau$, to extract the $\kappa \rm QC$ signal strength at 
the characteristic frequency $\kappa\omega_0$ completes the experimental protocol,
as elaborated 
in Sec.~\ref{sec:demod} below.

\section{Theoretical model}
\subsection{Hamiltonian}
Typical experiments to sense MQC signals are carried out at room temperatures
with atomic clouds of low densities $\rho\sim \SI{e7}-\SI{e11}{\per\centi\meter\cubed}$
\cite{lukas_bruder15,Bruder_2019}.
These are very dilute gases for which $\rho k_0^{-3}\ll 1$,
with
the atoms 
typically in the far-field of each other.
Given that collisional broadening (see, e.g.,~\cite{loudon})
in such ensembles is at least one order of
magnitude narrower than the natural line width~\cite{Bruder_phd},
we can neglect collisions between the atoms to a good approximation.

Our model considers $N$ identical, initially unexcited atoms at
random positions
$
  \mathbf{r}_\alpha(t)
  =
  {\bf r}_{\alpha0} + {\bf v}_\alpha t
$,
where ${\bf r}_{\alpha0}$ and ${\bf v}_\alpha$ are,
respectively, the initial coordinate and velocity of atom~$\alpha=1,\ldots,N$.
We assume a homogeneous and isotropic distribution of the initial coordinates
${\bf r}_{\alpha0}$ (which implements the above far-field configuration), 
while the Cartesian components of the velocities
${\bf v}_\alpha$ can be drawn from the Maxwell-Boltzmann distribution corresponding to
the temperature of the cloud.
At typical room temperatures the atomic velocity
is~$\sim \SI[per-mode=symbol]{e2}{\meter\per\second}$;
hence, the atomic momentum is much larger (for $^{87}$Rb atoms probed in
\cite{Bruder_2019}, by about three orders of magnitude) than the photon
momentum.
Under this condition we can neglect the recoil effect and treat the atomic
motion classically.

Then, the total Hamiltonian of the 
atomic ensemble interacting with an
electromagnetic bath represented by an infinite set of quantized harmonic
oscillators in their vacuum state, and with the classical, near-resonant pump
and probe 
pulses \eref{laser} reads
\begin{equation}
  H
  =
  H_0
  + H_\mathrm{F}
  + H_{\mathrm{AF}}
  + H_{\mathrm{AL}}
  ,
  \label{hamiltonian}
\end{equation}
where
\begin{itemize}
  \item
    $H_0 = \sum_{\alpha=1}^{N} \hbar\omega_0{\bf D}_\alpha^\dagger\cdot{\bf D}_\alpha$
    defines the free atoms' Hamiltonian, with 
    $\omega_0\approx\omega_{\rmL}$, and $\Du_\alpha$ ($\Dd_\alpha$) the atomic
    raising (lowering) operators incorporating arbitrary degeneracies of the ground
    and excited state sublevels.
  \item
    $H_\mathrm{F} = \sum_{{\bf k} s} \hbar \omega_ka^\dagger_{{\bf k}s}a_{{\bf k}s}$
    describes the contribution from the quantized, free electromagnetic environment,
    with $a^\dagger_{{\bf k}s}$ ($a_{{\bf k}s}$) the creation (annihilation)
    operator of a photon with wave vector~${\bf k}$ and polarization~$s$.
  \item $H_\mathrm{AL}$ and $H_\mathrm{AF}$ mediate the coupling between the
    atomic degrees of freedom and, respectively, the injected laser field and the
    quantized field.
    We employ the electric dipole approximation
    to write the Hamiltonians~$H_{\mathrm{AL}}$ and~$H_{\mathrm{AF}}$ as
    \begin{eqnarray}
      H_{\mathrm{AL}}
      &=&
      {-d} \sum_{\alpha=1}^N
        {\bf D}_\alpha^\dagger
        \cdot
        {\bf E}_{\rmL}({\bf r}_{\alpha}(t), t)
        \mathinner{
          \rme^{-\rmi 
             \omega_{\rmL} t
      }
        }
        + {\rm h.c.}
      ,
      \label{H_AL}
      \\
      H_{\mathrm{AF}}
      &=&
      {-d} \sum_{\alpha=1}^N
        {\bf D}_\alpha^\dagger
        \cdot
        \Bigl(
          \mathbf{E}^{(+)}(\mathbf{r}_\alpha(t))
          +
          \mathbf{E}^{(-)}(\mathbf{r}_\alpha(t))
        \Bigr)
        + {\rm h.c.}
      ,
      \label{H_AF}
    \label{H_int}
    \end{eqnarray}
   where $d$ is the atomic dipole matrix element \cite{loudon},
    which is assumed to be real, without loss of generality,
    ${\mathbf E}_{\rmL}(\mathbf{r}_\alpha(t),t)$
    is given by~\eref{laser} and
    ${\bf E}^{(+/-)}({\bf r}_\alpha(t))$
    is the positive/negative-frequency part of the quantized field given by
    \begin{equation}
      {\bf E}^{(+)}({\bf r})
      =
      \rmi \sum_{{\bf k}s}
      \left(
        \frac{\hbar \omega_k}{2\epsilon_0V}
      \right)^{1/2}
      \Eps_{{\bf k}s} a_{{\bf k}s}
      \rme^{\rmi {\bf k}\cdot{\bf r}}
      ,
      \qquad
      {\bf E}^{(-)}
      =
      \Bigl({\bf E}^{(+)}\Bigr)^\dagger,
    \end{equation}
    with $\epsilon_0$ the free space permittivity, $V$ the quantization volume, and
    $\Eps_{{\bf k}s}$ the polarization vector. 
    In Eq.~\eref{H_AL} for $H_{\mathrm{AL}}$, we have additionally applied the rotating-wave
    (resonant) approximation, whereas counter-rotating terms must be kept in
    Eq.~\eref{H_AF} for $H_{\mathrm{AF}}$ to obtain correct collective
    level shifts~\cite{Milonni74,PhysRevA.51.3128,trippenbach92}.
\end{itemize}


MQC signals were observed \cite{Bruder_2019} in $^{87}$Rb
vapours, with the probed transitions endowed with fine and hyperfine
structures, leading to rich multi-component features of the signals.
We here ignore these features and specialize to atoms 
equipped with a
$\JG = 0 \rightarrow \JE = 1$
transition, which involves a non-degenerate ground
state and three sublevels in the excited state (see \fref{fig:setup}).
While this simplifies our model with respect to the $^{87}$Rb experiment \cite{Bruder_2019}, it still incorporates the 
experimentally important excited state degeneracy 
(a feature already employed earlier \cite{shatokhin2005,shatokhin2006,ketterer2014} to account for the relevant optical transition $^1S_0\rightarrow\, ^1P_1$ in experiments on 
the coherent  backscattering of light by cold strontium atoms~\cite{bidel02}).
A crucial consequence thereof 
is the presence of
levels that are not driven by the polarized laser fields, but still can 
be populated by the dipole-dipole interaction.
Such processes can be identified using different pump-probe polarizations and observation directions
~\cite{Bruder_2019}. 
Unlike previous treatments
based on two-level atoms without any internal
degeneracy~\cite{Dai_2012,Mukamel_2016,Li_2017,Gao:16,PhysRevLett.120.233401,Yu:19},
our model is adequate to capture the consequences of this redistribution of
energy between angular momentum states.

The atomic lowering operator associated with a
$\JG = 0 \rightarrow \JE = 1$
transition is given by
\begin{equation}
  \Dd_\alpha
  =
  -\uvec{e}_{-1}\sigma_{12}^\alpha
  +\uvec{e}_0\sigma_{13}^\alpha
  -\uvec{e}_{+1}\sigma_{14}^\alpha,
  \label{dipole_operator}
\end{equation}
where $\sigma_{lk}^\alpha=\ket{l}_\alpha\bra{k}_\alpha$, the excited state
sublevels $\ket{2}$, $\ket{3}$, $\ket{4}$ (see \fref{fig:setup})
have magnetic quantum numbers $-1,0,+1$, and couple
to field polarizations
${\uvec{e}_{-1} = (\uvec{x}- \rmi \uvec{y})/\sqrt{2}}$,
${\uvec{e}_0 = \uvec{z}}$, and 
${\uvec{e}_{+1} = -(\uvec{x}+ \rmi \uvec{y})/\sqrt{2}}$, \footnote{Equivalently, we denote the unit polarization vectors in the Cartesian basis by $\hat{\bf e}_x$, $\hat{\bf e}_y$, and 
$\hat{\bf e}_z=\uvec{e}_0=\uvec{z}$.} respectively.

Note that, in equation \eref{hamiltonian}, we omit the kinetic energy of the atomic centre-of-mass motion, as well as atomic contact
interaction terms \cite{cohen-tannoudji_photons},
which must be taken into account in dense atomic systems, where
$\rho k_0^{-3}\gtrsim 1$,
but can be safely neglected in dilute ensembles \cite{Kupriyanov:2017oq}.

\subsection{Fluorescence intensity}
The experimentally detected total signal is expressed through the quantum-mechanical expectation value of the
intensity operator, averaged over atomic configurations,
\begin{equation}
  I_{\uvec{k}}(t)
  =
  \ConfAvg[\Big]{
    \bigl\langle
      \mathbf{E}^{(-)}_{\uvec{k}}(t) \cdot \mathbf{E}^{(+)}_{\uvec{k}}(t)
    \bigr\rangle
  }\,,
  \label{int}
\end{equation}
where
$\la\ldots\ra = \Trace \{ \ldots \varrho(0) \}$.
The initial density operator~$\varrho(0)$ of the atom-field system factorises
into the ground state of the atoms and the vacuum state of the field.
The positive frequency part~${\mathbf E}^{(+)}_{\uvec{k}}(t)$ of the
operator for the electric far-field
can be expressed through the atomic lowering operator of the source atoms ~\cite{agarwal74,ames2021}\footnote{Strictly speaking, due to the 
coupling described by the Hamiltonian $H_\mathrm {AF}$ (see \eref{H_int}), any atomic and field operators are \emph{atom-field} operators at $t>0$.},
\begin{equation}
  {\mathbf E}^{(+)}_{\uvec{k}}(t)
  =
  f \sum_{\alpha=1}^N
  \Bigl\{
    \Dd_\alpha(t)
    -
    \uvec{k} \bigl[ \uvec{k}\cdot \Dd_\alpha(t) \bigr]
  \Bigr\}
  \rme^{-\rmi\vk\cdot\mathbf{r}_\alpha(t)}
  \, ,
  \label{Eplus}
\end{equation}
where $f=\omega_0^2d/(4\pi\epsilon_0c^2R_d)$, $c$ is the speed of light,
$R_d$ is the distance from the center of mass of the cloud to the detector, and
${\mathbf k}$ is the wave vector of the scattered light.
In writing \eref{Eplus}, we assumed that
$
  R_d\gg r_{\alpha\beta}(t)
  =
  | \mathbf{r}_\alpha(t) - \mathbf{r}_\beta(t) |
$
for any $\alpha$, $\beta$.
Finally, $\ConfAvg{\ldots}$~stands for the configuration (or disorder)
average.
It results from the random initial positions of the atoms within
the cloud, and from their thermal motion during the fluorescence detection time $t_{\rm fl}$.
We will treat the thermal motion effectively, through configuration
averaging which involves an integration over the length and direction of the
vectors~$\mathbf{r}_{\alpha \beta}$ connecting pairs of atoms (see \sref{sec:disorder} for more details), with 
mean interatomic distance 
$\bar{r}$.\footnote{%
  $\bar{r}\simeq 0.554 \rho^{-1/3}$ \cite{PhysRevA.49.146}, which for a particle density $\rho\simeq 10^7-10^{11}$ cm$^{-3}$ \cite{lukas_bruder15,Bruder_2019} is about $\SI{100}-\SI{1}{\micro\meter}$.
}

Plugging~\eref{Eplus} into~\eref{int}, with atomic levels
from~\eref{dipole_operator}, and $\mathbf{k}=k\uvec{x}$, we obtain
\begin{equation}
  I_{\uvec{x}}
  =
  \ConfAvg[\bigg]{
    f^2
    \sum_{\mathclap{\alpha, \beta=1}}^N
    \rme^{\rmi k\uvec{x}\cdot\mathbf{r}_{\alpha\beta}}
    \Bigl(
      \la\sigma_{\rmz1}^\alpha\sigma_{1\rmz}^\beta\ra
      +
      \la\sigma_{\rmy1}^\alpha\sigma_{1\rmy}^\beta\ra
    \Bigr)
  }
  \label{numer_I}
\end{equation}
and $I_{\uvec{y}}$ follows upon replacing
$\uvec{x}\rightarrow \uvec{y}$,
$\rmy \rightarrow \rmx$
in \eref{numer_I}.
Note that, in the above equation, we dropped the time argument and introduced a
new (degenerate) atomic excited manifold basis,
\begin{equation}
\ket{\rmx} = (\ket{4}-\ket{2})/\sqrt{2},\quad 
\ket{\rmy} = (\ket{4}+\ket{2})/(\sqrt{2}\rmi), \quad
\ket{\rmz} = \ket{3}.
\label{xyz}
\end{equation}

\subsection{Master equation}
Evaluation of the fluorescence intensity \eref{numer_I} includes
certain time-dependent atomic dipole correlators like $\la\sigma_{\rmz1}^\alpha\sigma_{1\rmz}^\beta\ra$.
For immobile atoms it is technically involved, though fundamentally straightforward, to derive a Lehmberg-type master
equation for an arbitrary atomic operator \cite{PhysRevA.2.883} (or,
equivalently, for the atomic density operator \cite{agarwal74}) under
the standard Born and Markov approximations \cite{breuer_book}.
Such a master equation contains a Liouvillian describing the dipole-dipole
coupling, which is a function of the distance $r_{\alpha\beta}$ between the
interacting particles.
In a thermal ensemble, the interatomic distance is time-dependent, such that employing the standard approximations in this regime requires justification.
As it turns out, the Born and Markov approximations may also prevail for
moving atoms---provided that the Doppler shift $k_0v_{\alpha}$ ($v_\alpha = |{\bf v}_\alpha|$ and
$k_0=\omega_0/c$) amounts to a negligibly small phase shift $k_0v_{\alpha} r_{\alpha\beta}(t)/c\ll 1$ during the propagation time $r_{\alpha\beta}(t)/c$ 
of the interaction between the atoms 
\cite{trippenbach92}. In other words, the atoms can be considered ``frozen'' on the time scale  $r_{\alpha\beta}(t)/c$.
This condition is well satisfied, for example, for $r_{\alpha\beta}(t)=\bar{r}\sim  \SI{10}{\micro\meter}$ and a Doppler shift
$k_0v_\alpha\sim \SI{600}{\mega\hertz}$, since
$k_0v_\alpha r_{\alpha\beta}(t)/c\sim 2\times 10^{-5}$.
We therefore obtain the same form of the dipole-dipole interaction as for fixed
atoms, where $r_{\alpha\beta}$ is substituted by $r_{\alpha\beta}(t)$
\cite{trippenbach92}.

Thus, an arbitrary time-dependent atomic dipole correlator can be obtained from the
quantum master equation \cite{ames2021,PhysRevA.2.883} (written in the interaction picture with respect to the free-atom Hamiltonian $H_0$\footnote{The atomic correlators in equation \eref{numer_I} are invariant under this tranformation.}): 
\begin{equation}
    \dot Q
  =(\LiouL+\LiouG+\LiouInt)Q,
  \label{meq}
\end{equation}
where $Q$ is an arbitrary \emph{atomic} operator (on the space of $N$ atoms) \emph{averaged over the free field} subsystem, 
$\LiouL=\sum_{\alpha=1}^N\LiouL^\alpha$, with $\LiouL^\alpha$ 
the laser-atom interaction Liouvillian; $\LiouG=\sum_{\alpha=1}^N\LiouG^\alpha$, with $\LiouG^\alpha$ 
the Lindblad superoperator describing the atomic relaxation due to the electromagnetic bath; and $\LiouInt=\sum_{\alpha\neq \beta=1}^N\Liou_{\alpha\beta}$, 
with $\Liou_{\alpha\beta}$ 
the Liouvillian describing the dipole-dipole interactions mediated by the 
(common) electromagnetic bath. 
The Liouvillians read \cite{shatokhin2005,shatokhin2006,PhysRevA.51.3128,ames2021}
\begin{eqnarray}
  \LiouL^\alpha
  Q
  &=&
  -\frac{\rmi d}{\hbar}
  \bigl[
    {\mathbf D}_\alpha^\dagger
    \cdot
    {\mathbf E}_{\rmL}(\mathbf{r}_{\alpha 0},t)\rme^{-\rmi \delta_\alpha t}
    +
    {\mathbf D}_\alpha
    \cdot
    {\mathbf E}^*_{\rmL}(\mathbf{r}_{\alpha 0},t)\rme^{\rmi \delta_\alpha t}
    ,
    Q
  \bigr]
  ,
  \label{L_a}
  \\
  \LiouG^\alpha
  Q
  &=&
  \frac{\gamma}{2}
  \Bigl(
    \Du_\alpha
    \cdot
    \bigl[Q,\Dd_\alpha\bigr]
    +
    \bigl[\Du_\alpha,Q\bigr]
    \cdot
    \Dd_\alpha
  \Bigr)
  \label{L_g}
  ,
\end{eqnarray}
with
\begin{equation}
	\delta_\alpha=\omega_{\rmL}-\omega_0-\mathbf{k}_{\rmL}\cdot\mathbf{v}_\alpha
	\label{detuning}
\end{equation}
 the laser-atom detuning incorporating the Doppler shift of the atomic transition frequency,
$\gamma$ the spontaneous decay rate (identical for all excited state levels), and
\begin{equation}
  \Liou_{\alpha\beta}
  Q
  =
  \Du_\alpha
  \! \cdot \!
  \Tensor{T}(k_0r_{\alpha\beta}(t),\uvec{n})
  \! \cdot \!
  \bigl[Q, \Dd_\beta\bigr]
  +
  \bigl[\Du_\beta, Q\bigr]
  \! \cdot \!
  \Tensor{T}{}^*(k_0r_{\alpha\beta}(t),\uvec{n})
  \! \cdot \!
  \Dd_\alpha
  \, ,
  \label{L_ab}
\end{equation}
with the dipole-dipole interaction tensor
$\Tensor{T}(k_0r_{\alpha\beta}(t),\uvec{n})$ which, with 
$\uvec{n}=\mathbf{r}_{\alpha\beta}/r_{\alpha\beta}=(\sin\theta\cos\phi,\sin\theta\sin\phi,\cos\theta)$ and $\xi(t) =k_0r_{\alpha\beta}(t)$, reads
\begin{equation}
\fl
  \Tensor{T} (\xi(t), \uvec{n})
  =
  \frac{3\gamma}{4} \rme^{-\rmi\xi(t)}
  \left[
    \frac{\rmi}{\xi(t)}
    \Bigl( \Id - \uvec{n}\uvec{n} \Bigr)
    +
    \left(
      \frac{1}{\xi(t)^2}
      -
      \frac{\rmi}{\xi(t)^3}
    \right)
    \Bigl( \Id - 3 \uvec{n}\uvec{n} \Bigr)
  \right] .
  \label{tensor_T}
\end{equation}
In this equation, $\Id$ is a $3\times 3$ unit matrix and $\uvec{n}\uvec{n}$ is a
$3\times 3$ dyadic. Since we assume linear polarization of the incoming laser pulses, it is most convenient to express the matrix elements of $\Tensor{T}(\xi(t), \uvec{n})$ in the Cartesian basis, 
$
\Tensor{T}_{kl} 
=
\uvec{e}_k
\cdot
\Tensor{T} 
\cdot
\uvec{e}_{l}
$, 
$k, l \in \{ x, y, z \}$.

\subsection{Near- versus far-field form of the dipole-dipole interaction}
\label{sec:asymp_dd}
Note that
the near-field term in equation \eref{tensor_T}, scaling as
$\xi^{-3}$ (we drop for brevity the time argument),
is the
$\xi\ll 1$ asymptote of the dipole-dipole coupling
tensor~\eref{tensor_T}~\cite{PhysRevA.2.883}, and is associated with the electrostatic interaction.
It is this form of the interaction that was employed in all previous
theoretical treatments dealing with MQC signals in dilute atomic
gases~\cite{Dai_2012,Li_2017,Gao:16,PhysRevLett.120.233401,Yu:19}, where atomic angular momentum was also ignored. 

To identify the associated form of dipolar interactions, we consider a simple two-level atomic structure rather than our $\JG=0 \rightarrow \JE=1$ transition, i.e.\ we keep only level $|3\rangle$ in the excited state manifold. The tensor \eref{tensor_T} then reduces to its matrix
element $\Tensor{T}_{zz}$ corresponding to the magnetic number zero, which, in the near-field limit $\propto \xi^{-3}$, reads
\begin{equation}
  \Tensor{T}_{zz}
  	=- 
  	\frac{3 \gamma \rmi }{4}\frac{1 - 3 \cos^2\theta}{\xi^3}.
  \label{element_T00}
\end{equation}
Recalling the definition
$\gamma=4\omega_0^3d^2/(4 \pi \epsilon_0 3\hbar c^2)$ \cite{loudon},
and introducing the dipole moment operators
$\mathbf{d} (\sigma_{13}+\sigma_{31})$
of the two-level systems, with
$\mathbf{d}=d \uvec{e}_z$,
equations~\eref{meq}, \eref{L_ab} and \eref{element_T00} 
lead to the following 
Heisenberg equation of motion for two atoms $\alpha$ and $\beta$:
\begin{equation}
 \dot{Q}=
  \frac{\rmi}{\hbar}
  \Biggl[
    \underbrace{
      \frac{1}{ 4 \pi \epsilon_0 }
      \frac{
        {\mathbf d}^2
        -
        3({\mathbf d}\cdot{\uvec{n}})^2
      }{
        r_{\alpha\beta}^3
      }
    }_{
      V_{\text{d-d}}
    }
      \sigma_{31}^\alpha \sigma_{13}^\beta
    ,
    Q
  \Biggr],
  \label{static}
\end{equation}
where we have identified the electrostatic dipole-dipole interaction~$V_{\text{d-d}}$.
We can see from equation \eref{static} that the interatomic potential~$V_{\text{d-d}}$ generates purely Hamiltonian dynamics, which does not affect the ground and doubly excited states $|1\rangle_\alpha|1\rangle_\beta$ and $|3\rangle_\alpha|3\rangle_\beta$, respectively, of the uncoupled system. It does, however, shift both uncoupled  single-excitation eigenstates $|3\rangle_\alpha|1\rangle_\beta$ and $|1\rangle_\alpha|3\rangle_\beta$.
In quantum electrodynamics, these shifts are interpreted as originating from
the exchange of virtual photons \cite{cohen-tannoudji_photons}.

In contrast to the electrostatic part \eref{static} of the interaction, its radiative part
accounts for the exchange of real, or transverse, photons  \cite{cohen-tannoudji_photons}, and 
 generates \emph{both},
collective shifts \emph{and}
collective decay of the atomic excited states, associated, respectively, with the tensors
$
  \Tensor{\Omega}(\xi, \uvec{n})
  =
  \Im \Tensor{T}(\xi, \uvec{n})
$
 and
$
  \Tensor{\Gamma}(\xi, \uvec{n})
  =
  \Re \Tensor{T}(\xi, \uvec{n})
$.
While $
\Tensor{\Gamma}(\xi, \uvec{n})=0$ in the near-field limit \eref{static}, we here are dealing with dilute gases characterized by interatomic distances that are much larger than the optical wavelength.
Consequently, in the far-field limit $\xi \gg 1$, the tensor $\Tensor{T}$
retains its real and imaginary parts, scaling as $\xi^{-1}$, by equation \eref{tensor_T}.
In that limit, 
  $\Tensor{T}(\xi,\uvec{n})
  \approx
  (3 \gamma / 4)
  g(\xi)
  [\Id - \uvec{n}\uvec{n}]
$,
with coupling parameter
$
  g(\xi)
  =
  \rmi\rme^{-\rmi\xi}/\xi
$,
$
  |g(\xi)| \ll 1
$,
and decay rates and level shifts can be deduced from
\begin{equation}
\label{tensor_G_Om}
  \Tensor{\Gamma}(\xi, \uvec{n})
=
  \frac{3 \gamma}{4}
  \frac{\sin\xi}{\xi}
  (\Id - \uvec{n}\uvec{n})
  , \qquad
  \Tensor{\Omega}(\xi, \uvec{n})
 =
  \frac{3 \gamma}{4}
  \frac{\cos\xi}{\xi}
  (\Id - \uvec{n}\uvec{n})
  ,
\end{equation}
which are both oscillating functions of $\xi$, with identical amplitude. This means that both terms contribute equally to the dynamics and, hence, the
non-Hermititian dissipative terms generated by~$\Tensor{\Gamma}(\xi, \uvec{n})$
cannot be neglected.

With the definition \eref{tensor_G_Om} of $\Tensor{\Gamma}$ and $\Tensor{\Omega}$, the interaction Liouvillian from \eref{L_ab} can be decomposed as ${\cal L}_{\alpha\beta}={\cal L}^{(1)}_{\alpha\beta}+{\cal L}^{(2)}_{\alpha\beta}$,
with
\begin{eqnarray}
{\cal L}^{(1)}_{\alpha\beta}Q
&=&
-\rmi \Du_\alpha
\! \cdot \!
(\Tensor{\Omega}-\rmi \Tensor{\Gamma})
\! \cdot \!
\Dd_\beta Q
+
\rmi Q \Du_\alpha
\! \cdot \!
(\Tensor{\Omega}+\rmi \Tensor{\Gamma})
\! \cdot \!
\Dd_\beta,\label{Lab1}\\
{\cal L}^{(2)}_{\alpha\beta}Q
&=&
2 \Du_\alpha
\! \cdot \!
\Tensor{\Gamma}
\! \cdot \! Q
\Dd_\beta.\label{Lab2}
\end{eqnarray}
This distinction between contributions to ${\cal L}_{\alpha\beta}$ will be essential for our interpretation of MQC spectra in section \ref{sec:spectra} below. 
Let us therefore describe their action on specific operators $Q$ in detail:
\begin{itemize}
  \item The operator part
    $(\Du_\alpha)_r(\Dd_\beta)_{j}= \sigma^\alpha_{r1} \sigma^\beta_{1j}$,
    ($r,j\in \{\rmx,\rmy,\rmz\}$)
    of the Liouvillian~${\cal L}^{(1)}_{\alpha\beta}$ (see~\eref{Lab1})
    transfers one excitation from atom~$\beta$ to atom~$\alpha$.
    When $(\Du_\alpha)_r(\Dd_\beta)_{j}$ acts on the projector
    $Q = \sigma^\alpha_{11}\sigma^\beta_{ff}$ 
    from the left (where ($f\in \{\rmx,\rmy,\rmz\}$),
    it transforms the latter into the operator
    $Q'=\delta_{j f}\sigma^\alpha_{r1}\sigma^\beta_{1f}$
    of electronic coherences of both atoms.
    This transformation amounts to the dynamical equation
    $\dot{Q}= \sum_{rj}(-\rmi \Tensor{\Omega}_{rj}-\Tensor{\Gamma}_{rj})Q^\prime$,
    which couples the exited state populations to the electronic coherences.
    The action of $\sigma^\alpha_{r1} \sigma^\beta_{1j}$ from the right can be
    described analogously.
  \item Since the raising and lowering operators of different atoms in
    ${\cal L}^{(2)}_{\alpha\beta}$
    (see~\eref{Lab2}) act from the left and right side on an atomic operator,
    this Liouvillian performs the transformation to a two-atom subspace with an
    extra excitation.
    As a result, an operator $Q'=\sigma^\alpha_{1j}\sigma^\beta_{r1}$ describing electronic coherences of both atoms is transformed into that of Zeeman coherences and excited state populations, $(\Du_\alpha)_{f}Q^\prime(\Dd_\beta)_{l}=\sigma^\alpha_{fj}\sigma^\beta_{rl}=Q''$ ($f,j,r,l\in \{\rmx,\rmy,\rmz\}$).
    We can write this as a dynamical equation,
    $\dot{Q'}=2\sum_{fl}\Tensor{\Gamma}_{fl}Q''$,
    stating that Zeeman coherences and excited state populations of two atoms
    (embodied in $Q^{\prime\prime}$) are transformed to their electronic
    coherences (embodied in $Q^\prime$) via a collective decay process.
\end{itemize}

\section{Analytical solution of the master equation}
\label{sec:solution}

\subsection{Separation of time scales}
\label{sec:separation}
Our approach to solve~\eref{meq}, with ingredients \eref{L_a}-\eref{tensor_T}, analytically 
is based on the
observation that the atomic dynamics during each experimental cycle (see  \fref{fig:timescales}) is characterized by four
different time scales (recall~\sref{sec:spectroscopy}).
The \emph{first}, ultrashort time scale is set by the laser pulse length $\sigma \sim \SI{50}{\femto\second}$ (which enters through \eref{laser}).
The interpulse delay $\tau \sim \SI{10}{ps}$, which parametrizes \eref{intensity}, determines the \emph{second},
short time scale.
After the second pulse, atomic fluorescence is collected until the end of the cycle, at time $\Tcyc$ (entering \eref{intensity}, via $\phi_1$ and $\phi_2$, through $\tau_m$), until all atoms
relax into their ground states by emission of a photon.
This stage of the fluorescence detection establishes the \emph{third} time scale, which typically
lasts for about four orders of magnitude longer than the short time scale---for
several microseconds.
Finally, within the long time scale, the typical thermal coherence time~$\tau_{\mathrm{th}}$
\cite{PhysRevLett.97.013004} of the order of $\SI{10}{\nano\second}$\footnote{%
  The thermal coherence time $\tau_{\mathrm{th}} = \lambda_0/v=2\pi/k_0v\approx\SI{10}{\nano\second}$ can be
  obtained from the Doppler shift  $k_0v\approx\SI{560}{\mega\Hz}$
  \cite{Bruder_2019}.
}
defines the \emph{fourth} time scale, which demarcates the regime of the atomic
dynamics beyond which the Doppler effect is not negligible.

Such separation of time scales motivates a simplified version of~\eref{meq}, which decomposes into terms relevant at distinct time scales.
We thus completely ignore the incoherent
dynamics generated by~$\LiouG$ and
$\Liou_\text{int} $
\label{Lint}
during that ultrashort time scale, while, on longer time scales, we solve~\eref{meq} including all
terms from~\eref{L_a} and \eref{L_ab} \emph{except} for the Liouvillians~$\LiouL$ which generate the dynamics induced by the (on these long time scales) ``kick''-like laser-atom interaction. The thus defined propagation over ultrashort and longer time scales is then interchanged twice, to cover a single experimental cycle of duration $\Tcyc$. This approach resembles a stroboscopic description of the dynamics familiar, e.g., from stroboscopic maps \cite{wimberger_book} or also from micromaser theory \cite{PhysRevA.52.2361}.

Furthermore, we need to blend in the atoms' thermal motion, which can be safely neglected on the ultrashort and short time scales. The atomic configuration set by $\mathbf{r}_{\alpha 0}$ is therefore considered frozen on these time scales.  On the long time scale, however, atomic motion does affect the dipole-dipole coupling
$\Liou_{\rm int}$ via the then time-dependent interatomic vectors
${
  \mathbf{r}_{\alpha\beta}(t)
  =
  \mathbf{r}_{\alpha 0}
  - \mathbf{r}_{\beta 0}
  + (\mathbf{v}_\alpha - \mathbf{v}_\beta) t
}$. 
This leads to rapidly oscillating, ~$r_{\alpha\beta}(t)$-dependent phase factors in certain terms of our solutions of \eref{meq}, which average out for many repetitions of experimental cycles of 
duration $\Tcyc$, each of them being much longer than the thermal coherence time $\tau_{\mathrm{th}}$. Therefore, we effectively account for the thermal atomic motion by averaging over the atomic configurations. 
  As a result, atomic dipole correlators in~\protect\eref{numer_I} with
  position-dependent phases vanish, regardless of the observation direction.
  In contrast, so-called incoherent terms \cite{shatokhin2006}, which do not exhibit such phases and correspond to ~$\alpha = \beta$ in equation~\protect\eref{numer_I}, prevail.
  We note that, although this reduces the observable to a sum of single-atom
  correlators, these are obtained from the solution of the master
  equation~\protect\eref{meq} for~$N$ dipole-dipole \emph{interacting} atoms.
  Therefore, through these interactions, the fluorescence intensity~\protect\eref{numer_I}
  still depends on the surrounding atoms.
  An analogous reasoning applies for the position-dependent phases in the coupling~\protect\eref{tensor_T} and in the pulses' envelope (see equation \protect\eref{laser}), as will be detailed in \protect\sref{sec:single_double}.

\subsection{Dynamics of independent atoms: Single-atom case}
\label{sec:nonint_single}
In the following, we present our derivation of
$I_{\uvec{k}}(t)$ from \eref{int}.
For noninteracting atoms (that is, $\Liou_{\rm int}=0$), it suffices
to find a solution of a single-atom equation, as we show in this subsection.
The dynamical behaviour of $N$ atoms is then given by the tensor product of
evolution laws for individual atoms, as we discuss in \ref{sec:nonint_multi}.

By equation \eref{meq}, an arbitrary single-atom operator $Q_\alpha$ obeys the equation of motion
\begin{equation}
  \dot Q_\alpha(t)
  =
  \left(
     \LiouL^\alpha(t) + \LiouG^\alpha
  \right) Q_\alpha(t)
  ,
  \label{meq1}
\end{equation}
where the Liouvillians on the right hand side of \eref{meq1} are
given by  \eref{L_a}, \eref{L_g}. 

\subsubsection{Ultrashort time scale --- atom-laser interaction.}
\label{sec:ultrashort-pulses}
As argued in \protect\sref{sec:separation}, during the ultrashort intervals of
non-vanishing
laser intensity around the pulses' arrival times~$t_j$,
dissipation processes generated by~$\LiouG^\alpha$ are insignificant.
Hence, we need to solve the following equation of motion generated by the Liouvillian $\LiouL^\alpha$:
\begin{equation}
\dot Q_\alpha(t)
=
\LiouL^\alpha(t) Q_\alpha(t)
=
- \frac{\rmi}{2}\Omega_{\rmL}(t)
[{\cal M}_j(t), Q_\alpha(t)]
,
\label{meq2}
\end{equation}
where $\Omega_{\rmL}(t)=2d{\cal E}_{\rmL}(t-t_j)/\hbar$ is the time-dependent
(real-valued) Rabi frequency, and
\begin{equation}
{\cal M}_j(t)
=
S^\dagger \rme^{\rmi \chi_j - \rmi \delta_\alpha t} 
+ S \rme^{-\rmi \chi_j + \rmi \delta_\alpha t}\,.
\label{def_M}
\end{equation}
This is a generalized atomic dipole operator (whose index $\alpha$ we drop for simplicity), with
$S^\dagger = {\bf D}^\dagger_\alpha \cdot \EpsL$,
$S = {\bf D}_\alpha\cdot \EpsL^*$
atomic raising and lowering operators,
according to~\eref{dipole_operator}, and the time-independent phase 
\begin{equation}
\chi_j 
=
{\bf k}_{\rmL} \cdot {\bf r}_{\alpha 0}
+ \omega_{\rmL} t_j
+ w_j \tau_m\, ,
\label{phase_phi_j}
\end{equation}
which includes the phase tag $w_j \tau_m$ proper, as well as the phases imprinted by
the carrier frequency according to 
\eref{laser}.
The laser-atom detuning $\delta_\alpha$
is given by equation \eref{detuning}.

Since the interaction with the Gaussian laser pulses 
occurs instantaneously in comparison to
all other time scales (see section \ref{sec:separation} and fig.~\ref{fig:timescales}), these 
can be modelled by delta pulses carrying the same energy
$E_{\rm pulse}$.
Such pulses are characterised solely by their area
\begin{equation}
  \vartheta
  =
  \int_{-\infty}^\infty \dif{t'}
    \Omega_{\rmL}(t')
  ,
  \label{pulse_area}
\end{equation}
which can be found using the relation
(see, for instance, \cite{siegman_lasers})
between the energy and the intensity of a pulse,
\begin{equation}
  E_{\rm pulse}
  =
  2\epsilon_0cw_0{\cal E}^2_0
  \int_{-\infty}^\infty \dif{t'}
    \rme^{-(t'/\sigma)^2}=2\epsilon_0cw_0{\cal E}_0^2\sigma\sqrt{\pi}\,  ,
  \label{Energy}
\end{equation}
with $w_0$ the laser beam's cross-section.
Solving~\eref{Energy} for the field amplitude ${\cal E}_0$, we obtain
$
  \vartheta
  =
  2d(\sigma c \mu_0 E_{\rm pulse}\sqrt{\pi}/w_0)^{1/2}/\hbar
$,
with ~$\mu_0$ the free space permeability.

Furthermore, under the delta pulse approximation, equation \eref{meq2} is simplified by replacing $\Omega_{\rmL}(t)$ by a properly weighted Dirac delta-function, and using the well-known formula (see, e.g.\ \cite{schiff}) $f(t)\delta(t-t_j)=f(t_j)\delta(t-t_j)$. As a result, 
we can replace ${\cal M}_j(t)\rightarrow {\cal M}_j(t_j)\equiv {\cal M}_j$, with
\begin{equation}
{\cal M}_j=S^\dagger \rme^{\rmi \varphi_j}+S \rme^{-\rmi \varphi_j },
\label{redef_M_j}
\end{equation}
where
\begin{equation}
\varphi_ j =
{\bf k}_{\rmL} \cdot {\bf r}_{\alpha 0}
+ \omega_0 t_j
+ w_j \tau_m 
+
\Delta_\alpha t_j,
\label{varphij}
\end{equation}
and $\Delta_\alpha={\bf k}_\rmL\cdot{\bf v}_\alpha$ is the Doppler shift of atom $\alpha$.
Now, the action of a delta kick at time $t_j$ on an atom can be expressed
through a unitary superoperator~$R^\alpha_j$ as
\begin{equation}
  Q_\alpha(t^+_j)
  =
  R^\alpha_j
  Q_\alpha(t_j^-)
  =
  \mathinner{
    \rme^{-\rmi\vartheta{\cal M}_j/2}
  }
  Q_\alpha(t_j^-)
  \mathinner{
    \rme^{\rmi\vartheta{\cal M}_j/2}
  }
  ,
  \label{kick}
\end{equation}
where $Q_\alpha(t_j^-)$ ($Q_\alpha(t_j^+)$) is 
the single-atom 
operator before (after) the kick.

We can deduce the transformation performed by $R^\alpha_j$ in an algebraic form.
To that end,  we notice that the operators $S^\dagger$ and $S$ on the right
hand side of equation \eref{def_M} are generalizations of the Pauli pseudo-spin
operators $\sigma_+$ and $\sigma_-$ to the case of degenerate excited state
sublevels.
Therefore, the algebraic properties of $S^\dagger$, $S$ are similar to those of
$\sigma_+$, $\sigma_-$.
To illustrate this by an example, let us consider laser polarizations
$\EpsL = \uvec{x}$ and
$\EpsL = \uvec{y}$, for which the operator $S$
reads, respectively (see also \eref{xyz}),
\begin{equation}
  \uvec{x}:
    S = \frac{\sigma_{14} - \sigma_{12}}{\sqrt{2}}
  ,\quad
  \uvec{y}:
    S = \rmi \frac{\sigma_{14} + \sigma_{12}}{\sqrt{2}},
  \label{lowering_xy}
\end{equation}
and the corresponding raising operators follow by Hermitian conjugation of \eref{lowering_xy}.
Using equations \eref{def_M} and \eref{lowering_xy}, we obtain
\begin{equation}
  {\cal M}_j^2
  =
  S^\dagger S + S S^\dagger,
\end{equation}
which is a projector onto the subspace of the laser-driven levels $\ket{1}$, $\ket{2}$, and $\ket{4}$.
This implies
\begin{equation}
  {\cal M}_j^{2k+1}
  =
  {\cal M}_j
  , \quad
  {\cal M}_j^{2k+2}
  =
  {\cal M}_j^2
  , \quad
  k=0,1,2,\ldots
  ,
\end{equation}
and, hence,
\begin{equation}
  \exp\Bigl(\pm\, \rmi \,\frac{\vartheta}{2} {\cal M}_j\Bigr)
  =
  \left(\Id - {\cal M}_j^2\right)
  + {\cal M}_j^2 \cos\frac{\vartheta}{2}
  \pm \rmi {\cal M}_j \sin\frac{\vartheta}{2}.
  \label{exp_M}
\end{equation}
The last two terms of the obtained result are formally equivalent to the known
operator identity for two-level systems (see, e.g.\ \cite{barnett_radmore}),
and describe an azimuthal rotation on the Bloch sphere by an angle~$\vartheta$.
In addition, the (trivial) evolution of laser-undriven levels (in the basis \eref{xyz}, and for $\EpsL = \uvec{x}$ ($\EpsL = \uvec{y}$), these are the 
levels $\ket{y}$ and $\ket{z}$ ($\ket{x}$ and $\ket{z}$)) 
is included in 
\eref{exp_M} via the projector $\Id-{\cal M}_j^2$.
With the aid of equations~\eref{exp_M} and~\eref{def_M}, we obtain \cite{beni_MSc} the following expansion for the superoperator $R_j^\alpha$ from~\eref{kick}:  
\begin{equation}
  R^\alpha_j
  =
  \sum_{p=-2}^2
    \rme^{\rmi p\varphi_j} R_j^{[p]},
  \label{phase_R1}
\end{equation}
with superoperators~$R_j^{[p]}$ independent of  $\varphi_j$, and defined by their action on a single-atom operator $Q_\alpha$:
\def \ProjUndrv {\mathinner{\Pi_\vartheta}}
\def \SPlus {\mathinner{S^\dagger_\vartheta}}
\def \SMinus {\mathinner{S_\vartheta}}
\begin{eqnarray}
  \label{Rj-split}
  \mathinner{R_j^{[0]}} Q_\alpha
  &=&
  \ProjUndrv Q_\alpha \ProjUndrv
  + \SPlus Q_\alpha \SMinus
  + \SMinus Q_\alpha \SPlus
  ,
  \nonumber \\
  \mathinner{R_j^{[+1]}} Q_\alpha
  &=&
  \rmi \left(\ProjUndrv Q_\alpha \SPlus-
    \SPlus Q_\alpha \ProjUndrv    
  \right)
  ,
  \nonumber \\
  \mathinner{R_j^{[-1]}} Q_\alpha
  &=&
  \rmi \left(\ProjUndrv Q_\alpha \SMinus -
    \SMinus Q_\alpha \ProjUndrv   
  \right)
  ,
  \\
  \mathinner{R_j^{[+2]}} Q_\alpha
  &=&
  \SPlus Q_\alpha \SPlus
  ,
  \nonumber \\
  \mathinner{R_j^{[-2]}} Q_\alpha
  &=&
  \SMinus Q_\alpha \SMinus
  ,
  \nonumber
\end{eqnarray}
where
\begin{equation}
  \ProjUndrv
  =
  \left(\Id - {\cal M}_j^2\right)
  + {\cal M}_j^2 \cos\frac{\vartheta}{2}
  ,
  \qquad
  \SMinus
  =
  S \sin\frac{\vartheta}{2}
  .
\end{equation}
Equations~\eref{phase_R1} and ~\eref{Rj-split} are crucial for the subsequent analysis of MQC signals in
\sref{sec:demod}, since
they
depend on the 
modulated phase $\varphi_j$ given by \eref{varphij}.
We note that the simple structure of the expansion~\eref{phase_R1} is a direct consequence of the RWA ~\cite{seidner1995} applied in the atom-laser
Hamiltonian $H_{\mathrm{AL}}$ as given in \eref{H_AL}.

While the decomposition \eref{phase_R1} is general, in our scenario the  atoms are initially unexcited. In this case, the atomic 
density operator after the first laser pulse does not feature factors $\rme^{\pm 2\rmi \varphi_1}$, which is important for the disorder average and 
signal demodulation (see \sref{sec:disorder} and \sref{sec:demod}). 
For the $\hat{\bf x}$-polarized pump pulse, using \eref{xyz}, \eref{lowering_xy}, \eref{phase_R1} and ~\eref{Rj-split}, we deduce
\begin{equation}
	R_{1,\hat{\bf x}}^\alpha\ketbra{1}{1}
	=\left(\cos\frac{\vartheta}{2}|1\rangle-\rmi \rme^{\rmi \varphi_1}\sin\frac{\vartheta}{2}|\rmx\rangle\right)\left(\cos\frac{\vartheta}{2}\langle 1|+\rmi \rme^{-\rmi \varphi_1}\sin\frac{\vartheta}{2}\langle \rmx|\right),
	\label{Rj-action-ground-state}
\end{equation}
where we additionally introduced the index $\hat{\bf x}$, to make the polarisation explicit, and 
$|\rmx\rangle=S^\dagger|1\rangle$.
The single-atom density operator after the pump pulse thus 
exhibits electronic coherences, modulated as $\rme^{\pm \rmi \varphi_1}$, and populations of the ground and excited state sublevels dependent on the pulse area. 

\begin{figure}
	\center
	\includegraphics[width=0.6 \columnwidth]{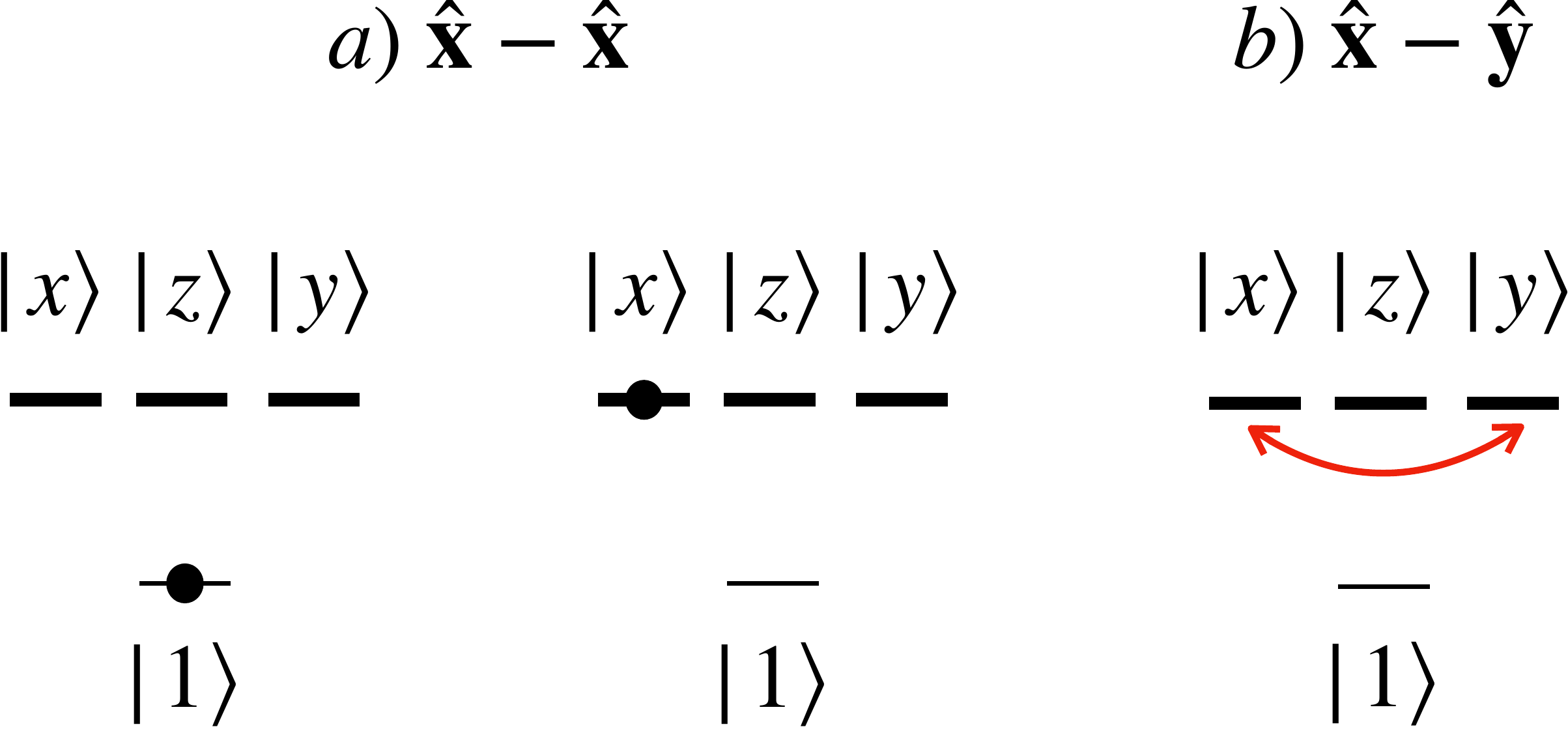}
	\caption{Graphical representation of the atomic internal variables (in the basis \eref{xyz}) that can yield MQC signals upon demodulation, as explained in \sref{sec:spectra}, for (a) $\uvec{x}$-$\uvec{x}$ and (b) $\uvec{x}$-$\uvec{y}$ pump-probe polarization. (a) (black dots) populations $\langle \sigma_{11}\rangle$ (left) and $\langle \sigma_{xx}\rangle$ (right) of states $|1\rangle$ and $|\rmx\rangle$, respectively, are modulated as $\cos(\varphi_1-\varphi_2)$,  (b) (red arc) Zeeman coherences $\langle \sigma_{xy}\rangle$ and $\langle \sigma_{yx}\rangle$ between the states $|x\rangle$ and $|y\rangle$  are modulated as $\exp(\pm\rmi(\varphi_1-\varphi_2))$.}
	\label{fig:coherences}
\end{figure}

In \sref{sec:single_double} below, by combining delta-pulse excitations with the subsequent time evolution due to
$\LiouG + \LiouInt$, we obtain an analytical formula for the two-atom
state after the second pulse.
In fact, for typical interpulse delays
$\tau\sim \SI{10}{\pico\second}$, the dynamics generated by the Liouvillians $\LiouG$ and  $\LiouInt$ can be neglected to a good approximation. 
With this in mind, we can 
derive explicit expressions for the states $|\psi_{1}\rangle\langle \psi_{1}|:=R^\alpha_{2,\hat{\bf x}}R^\alpha_{1,\hat{\bf x}}|1\rangle\langle 1|$ and $|\psi_{2}\rangle\langle \psi_{2}|:=R^\alpha_{1,\hat{\bf x}}R^\alpha_{2,\hat{\bf y}}|1\rangle \langle 1|$ of independent (since we ignore $\LiouInt$) atoms after a sequence of two $\uvec{x}$-$\uvec{x}$- and $\uvec{x}$-$\uvec{y}$-polarized pulses, respectively.  From equations \eref{kick}, \eref{exp_M}, and \eref{Rj-action-ground-state} we obtain 
\begin{eqnarray}
|\psi_{1}\rangle 
&=& \Bigl(\cos^2\frac{\vartheta}{2}+\rme^{\rmi(\varphi_1-\varphi_2)}\sin^2\frac{\vartheta}{2}\Bigr)|1\rangle
-\rmi(\rme^{\rmi \varphi_1}+\rme^{\rmi \varphi_2})\sin\frac{\vartheta}{2}\cos \frac{\vartheta}{2} |x\rangle,\label{psi_xx}\\
|\psi_{2}\rangle
&=&  \cos^2\frac{\vartheta}{2}|1\rangle-\rmi \rme^{\rmi \varphi_1}\sin\frac{\vartheta}{2}|x\rangle - \rmi \rme^{\rmi \varphi_2}\sin\frac{\vartheta}{2}\cos \frac{\vartheta}{2}|y\rangle.
\label{psi_xy}
\end{eqnarray}		
Let us now highlight those
matrix elements of 
$|\psi_{1}\rangle\langle \psi_{1}|$ and  $|\psi_{2}\rangle\langle \psi_{2}|$,
respectively, which 
can contribute to MQC signals upon demodulation.
Specifically, these 
are affected by the
phase modulation only through the difference $\varphi_1 - \varphi_2$.
It follows from \eref{psi_xx} that if 
both pulses have identical ($\uvec{x}$-$\uvec{x}$) linear
polarization, the second pulse transforms electronic coherences 
exhibited by \eref{Rj-action-ground-state} 
into
ground
and excited
state populations, $\langle \sigma_{11}\rangle$ and $\langle \sigma_{xx}\rangle$, respectively,
(see \fref{fig:coherences}(a)), both modulated as
$\cos(\varphi_1-\varphi_2)$.
In contrast, 
equation \eref{psi_xy} shows that, 
if the pulses are perpendicularly polarized ($\uvec{x}$-$\uvec{y}$), the 
Zeeman coherences (off-diagonal density matrix
elements in the excited state manifold) $\langle\sigma_{xy}\rangle$
and $\langle\sigma_{yx}\rangle$ are created (see \fref{fig:coherences}(b)), and 
modulated as
$\rme^{\pm \rmi (\varphi_1-\varphi_2)}$. Note that this phase modulation 
at short delays $\tau$ is independent of the fact that $\LiouG$ and $\LiouInt$ are 
dropped.

Before we proceed, let us notice that, so far, we 
described the action of the laser pulses on the atoms  \emph{non-perturbatively} in the pulse
area~$\vartheta$.
Strong pulses (e.g., $\vartheta \approx \pi$) induce a nonlinear response of the atoms ~\cite{mukamel_book,allen_eberly} and can be used as a control and diagnostic tool in electronic spectroscopy \cite{Gelin:2011dp,Binz:20}. Furthermore, we have shown that strong pulses substantially modify the MQC signals \cite{ames2021}.  However, in our present contribution we seek to elucidate the impact of the \emph{dipolar interactions} on these signals. Thus, we adhere to the
experimental conditions of~\cite{Bruder_2019} and
evaluate~\eref{Rj-split} for small~$\vartheta$, in the perturbative regime.

\subsubsection{Beyond the ultrashort time scale.}
\label{sec:single-atom-free-evolution}
We now derive from \eref{meq1} the time evolution generated by the Liouvillian
$\LiouG^\alpha$ as given by \eref{L_g}:
%
\begin{equation}
  \dot Q_\alpha
  =
 \LiouG^\alpha Q_\alpha
  ,
  \label{meq-free}
\end{equation}
with the formal solution (for simplicity, we take $t_0=0$)
\begin{equation}
  Q_\alpha(t)
  =
  \exp\bigl(
     \LiouG^\alpha  t
  \bigr) Q_\alpha(0)
  .
  \label{meq-free-exp}
\end{equation}
The explicit solution is derived in~\ref{app:solve-free-atom}, with the result
\begin{eqnarray}
  Q_\alpha(t)
  &=&
  \ProjE Q_\alpha(0) \ProjG
  \mathinner{
    \rme^{- \gamma t/2}
  }
  +
  \ProjG Q_\alpha(0) \ProjE
  \mathinner{
    \rme^{- \gamma t/2}
  }
  \nonumber \\
  &&+
  \ProjG Q_\alpha(0) \ProjG
  +
  \ProjE Q_\alpha(0) \ProjE
  \mathinner{
    \rme^{-\gamma t}
  }
  +
  {\bf D}_\alpha^\dagger\cdot Q_\alpha(0){\bf D}_\alpha
  \left( 1 - \rme^{-\gamma t} \right)
  ,
  \label{Qtsol}
\end{eqnarray}
where $\ProjE={\bf D}^\dagger_\alpha\cdot {\bf D}_\alpha$ is the projector on the
subspace of excited states, and $\ProjG=\Id -\ProjE$.
\Eref{Qtsol} has a transparent interpretation: the first line governs
the exponential decay of electronic coherences with rate~$\gamma/2$, while the
remaining terms ensure trace conservation and the decay of excited state
populations and Zeeman coherences
with rate~$\gamma$.


\subsection{Dynamics of independent atoms: Multi-atom case}
\label{sec:nonint_multi}
An arbitrary operator $Q(t)$ of $N$ atoms can now be obtained by combining single-atom operators $Q_\alpha$ in a tensor product, $Q(t)=\otimes_{\alpha=1}^NQ_\alpha(t)$.
Consistently, the evolution of a composite operator $Q(t)$ in the case of
non-interacting atoms is governed by superoperators
\begin{equation}
  R_j^{\rmL}
  =
  \bigotimes_{\alpha=1}^N R^\alpha_j
  , \quad
  \exp\left( \Liou_\gamma t \right)
  =
  \bigotimes_{\alpha=1}^N
  \exp\left( \Liou^\alpha_\gamma t \right),
  \label{def_A_G_0}
\end{equation}
where $\LiouG = \sum_{\alpha = 1}^N \LiouG^\alpha$,
the action of $R^\alpha_j$ is defined by equation \eref{Rj-split}, and that of
$\LiouG^\alpha$ by \eref{Qtsol}. The superoperator
$R^{\rmL}_j$ describes the coherent interaction of the individual atoms with the ultrashort
incoming pulses, whereas $\exp(\LiouG t)$ governs their incoherent dynamics 
when the laser field is switched off. 

We stress that, by \eref{Qtsol}, 
this dynamics amounts to an \emph{exponential decay} of all atomic variables in the multi-atom excited state manifold (that is, all excited state populations, 
as well as Zeeman and electronic coherences), while the collective ground state of all atoms is unaffected by $\exp(\LiouG t)$.

Finally, we introduce the resolvent superoperator ${\cal G}_\gamma (z)$, $z\in \mathbb{C}$, which
is the Laplace image of the evolution superoperator $\exp(\Liou_\gamma t)$,
\begin{equation}
  {\cal G}_\gamma(z)
  =
  \int_0^\infty \dif{t}
    \rme^{-z t} \rme^{\LiouG t}
  =
  \frac{1}{z - \LiouG}
  ,
  \label{G_0}
\end{equation}
and 
will be used in the next subsection, where we present 
the case of interacting atoms.

\subsection{Dynamics of the dipole-dipole interacting atoms}
\label{sec:d_d}

\subsubsection{Incorporation of \texorpdfstring{$\Liou_{\rm int}$}{the interaction}.}
As discussed in section \ref{sec:separation}, the dipole-dipole interaction is relevant 
on short as well as on long time scales.
Due to the weakness of the coupling parameter~$|g(\xi)|$
in the far field (see \sref{sec:asymp_dd}), the coupling~$\Liou_{\rm int}$
can be readily included perturbatively into our solutions~\eref{def_A_G_0}
for independent atoms. 
Note that this treatment is 
distinct from the traditional one~\cite{Dai_2012,Li_2017,Gao:16,PhysRevLett.120.233401,Yu:19}, where the dipole-dipole interactions are accounted for non-perturbatively,
via transformation to the exciton basis~\cite{mukamel_book}.

The starting point for the perturbative expansion in~$\Liou_{\rm int}$
is the equation of motion for an $N$-atom operator $Q$, recall \eref{meq} and fig.~\ref{fig:timescales},
\begin{equation}
  \dot{Q}
  =
    \LiouG Q
    +
    \Liou_{\rm int} Q
,
  \label{matr_L0_V}
\end{equation}
valid on short and long time scales.
Solving  \eref{matr_L0_V} perturbatively in $\Liou_{\rm int}$, by the Laplace transform method \cite{carrier_complex}, we obtain a series expansion
\begin{equation}
  \tilde{Q}(z)
  =
  \sum_{n=0}^\infty
    {\cal G}_\gamma(z)
    \Bigl[
      \Liou_{\rm int} {\cal G}_\gamma(z)
    \Bigr]^n
    Q(t_0)
  ,
  \label{series_Qz}
\end{equation}
where 
$
  \tilde{Q}(z)
  =
  \int_{t_0}^\infty \dif{t'}
    \exp(-z t') Q(t')
$, and $Q(t_0)$ is an operator which encodes the initial condition, to be specified below. 
The time evolution of $Q(t)$ then follows from \eref{series_Qz} via the \emph{convolution} theorem for Laplace transforms \cite{carrier_complex}.

\subsubsection{Combining piecewise solutions.}
\label{sec:combine}
Equation \eref{matr_L0_V} holds during the (short) interpulse delay $\tau$, and during the (long) fluorescence collection time $t_\mathrm{fl}$
(see \fref{fig:timescales}).
Therefore, when solving \eref{matr_L0_V}
with the aid of the
Laplace transform, we introduce distinct 
variables for each interval,
$z_1$ and $z_2$, respectively.
During the short interpulse delay, the time evolution is given by the inverse
Laplace transform
\begin{equation}
  Q(\tau)
  =
  \frac{1}{2 \pi \rmi}
  \int_{z_0 - \rmi \infty}^{z_0 + \rmi \infty} \dif{z_1}
    \rme^{z_1 \tau} \tilde{Q}(z_1),
  \label{inverseL}
\end{equation}
with $z_0 \in \mathbb{R}$ chosen such that $z_1$ lies in the half-plane of convergence \cite{carrier_complex},
and $\tilde{Q}(z_1)$ is given by \eref{series_Qz}.
The evolution during the long time interval is once again defined by~\eref{inverseL},
with the same $z_0$, and upon the replacements $z_1\rightarrow z_2$,
$\tau\rightarrow t_{\rm fl}$.

The dynamics of the atomic operators during the ultrashort intervals can be
absorbed in the Laplace transform solutions $\tilde{Q}(z_1)$ and $\tilde{Q}(z_2)$,
through the initial conditions specified by $Q(t_0)$ in~\eref{series_Qz}.
The short interval $\tau$ starts after the pump pulse at time $t_1$.
Hence, by \eref{kick}, we must set $Q(t_0)\equiv Q(t_1^+)=R_1^{\rmL}Q(t_1^-)$
in~\eref{series_Qz}, where $Q(t_1^-)=Q(0)$, i.e., it describes atomic operators
before any interaction took place.
Likewise, for the long time interval starting after the probe pulse, we set
$Q(t_0) \equiv Q(t_2^+) = R_2^{\rmL}Q(t_2^-)$
in~\eref{series_Qz}, with $Q(t_2^-)$
given by the inverse Laplace transform~\eref{inverseL} with~$t_2^-=\tau$. 
Thus, through the initial conditions $Q(t_2^+)$, a solution during the long time
interval includes the action of two laser pulses and the evolution during the
short time interval.
Combination of piecewise solutions, each of which is valid on distinct
time intervals, finally provides the evolution of an arbitrary atomic operator
throughout the entire excitation-emission cycle.

\subsubsection{Single and double scattering contributions.}
\label{sec:single_double}
The formal solution \eref{series_Qz} applies for an arbitrary number $N$ of atoms, though
we focus on the case $N=2$ in the following. 
At a first glance, this may appear a gross oversimplification of the experimentally studied 
physical system, including many atomic scatterers. 
Note, however, that we are dealing with a {\it dilute thermal} gas, whose fluorescence signal can be oftentimes inferred from the results for $N=1$ \cite{Mukamel_2016}. Yet, as we show below, one cannot measure a double quantum coherence signal by monitoring fluorescence of one, nor of two {\it non-interacting} atoms. Nor need we consider more than two atoms exchanging photons. Indeed,
the atoms are in the far field of, and detuned from each other. The latter condition stems from the motion-induced Doppler shifts being much larger than the natural linewidth. The photon \emph{scattering mean free path} \cite{guerin2017} in such medium can be estimated as\footnote{We here assume a fine transition of $^{87}$Rb atoms, with $\lambda_0=\SI{790}{\nano\meter}$, $\gamma=2\pi\times\SI{6}{\mega\hertz}$ \cite{steckRubidium87Line2015}, and typical experimental 
\cite{Bruder_2019} values of 
	the root-mean-square Doppler shift $\bar{\Delta}=\SI{560}{\mega\hertz}$, and $\rho=\SI{e8}{cm^{-3}}$.}
\begin{equation}
\ell_{\rm sc}=\frac{1}{\rho\bar{\sigma}_\mathrm{sc}}\approx \SI{88}{\meter}, 
\label{ell_sc}
\end{equation}
with $\bar{\sigma}_\mathrm{sc}$ the mean scattering cross-section (see \ref{sec:ell_sc}).
Therefore, a thermal atomic vapour of a reasonable size (say, a spherical cloud with a diameter $D$ of a few centimetres) is characterized by a very low \emph{optical thickness} \cite{Kupriyanov:2017oq} $b=D/\ell_{\rm sc}\ll 1$, such that long scattering sequences of photons, mediating interactions between many particles, are highly improbable. 
Consistently, single and double scattering contributions
do capture the essential physics in optically thin ensembles of scatterers \cite{jonckheere2000}, and 
we restrict our perturbative expansion \eref{series_Qz} to second order in~$\Liou_{\rm int}$.
This is the lowest-order contribution which survives the disorder average,
and it accounts for the double scattering\footnote{In section \sref{sec:spectra}, we consider in more detail the role played by the contributions ${\cal L}_{\alpha\beta}^{(1)}$ and ${\cal L}_{\alpha\beta}^{(2)}$ to ${\cal L}_{\alpha\beta}$ (see equations \eref{Lab1} and \eref{Lab2}) for the emergence of MQC signals. Until then, by `double scattering' we indicate all contributions surviving the disorder average that are second order in ${\cal L}_{\rm int}$.} of a single photon
\cite{ketterer2014}.  Thereby, we ignore 
{\it recurrent} scattering \cite{PhysRevB.50.16729}, where a photon bounces back and forth
between the atoms, and which corresponds to a higher-order expansion in~$\Liou_{\rm int}$. Neglecting such processes is fully justified in optically thin  media \cite{Kupriyanov:2017oq}.

Indeed, as we will see below, the above framework enables deducing 1QC and 2QC signals in good qualitative agreement with the experiment \cite{Bruder_2019},  and establishes the relevance of few-body systems in the characterization of MQC signals emitted by dilute thermal atomic ensembles. 

The single and double scattering contributions can be derived from the
two-dimensional Laplace transforms 
\begin{eqnarray}
\tilde{Q}^{[n]}(z_1,z_2,\varphi_1,\varphi_2)
&=&
\sum_{p=0}^n
{\cal G}_\gamma(z_2)
\Bigl[
\Liou_{\rm int}
{\cal G}_\gamma(z_2)
\Bigr]^{n - p}
\nonumber \\
&& \qquad \times
\mathinner{
	R_2^{\rmL}
}
{\cal G}_\gamma(z_1)
\Bigl[
\Liou_{\rm int}
{\cal G}_\gamma(z_1)
\Bigr]^p
\mathinner{
	R_1^{\rmL}
}
Q(0)
,
\label{Q2}
\end{eqnarray}
with $n=0,2$, which are obtained 
from equation \eref{series_Qz} and the contribution of 
piecewise solutions as described
in section \ref{sec:combine}. Note that
the phase tags (from equation \eref{varphij}) in the argument of \eref{Q2} are absorbed in $R_1^{\rmL}$ and $R_2^{\rmL}$, via \eref{phase_R1} and \eref{def_A_G_0}.

To infer the detected fluorescence intensity  from the operator equation
 \eref{Q2}, we need specific atomic correlation functions which correspond to $\alpha=\beta$ in equation \eref{numer_I} (see \sref{sec:separation}).
These are obtained by representing the operator equations in matrix form, via
the complete basis of the operator space for two atoms (see \ref{app:basis}), and subsequently evaluating 
the quantum mechanical expectation value, followed by the configuration average:
\begin{eqnarray}
\left\langle
\la\tilde{\bf Q}^{[n]}(z_1,z_2,\varphi_1,\varphi_2)\ra
\right\rangle_{\rm conf}
&=&
\sum_{p=0}^n
\ConfAvg[\Big]{
	{\bf G}_\gamma(z_2)
	\Bigl[
	{\bf V}
	{\bf G}_\gamma(z_2)
	\Bigr]^{n - p}
	\nonumber \\ && \hspace{-3em} \times
	{\bf R}_2^{\rmL}
	{\bf G}_\gamma(z_1)
	\Bigl[
	{\bf V}
	{\bf G}_\gamma(z_1)
	\Bigr]^p
	{\bf R}_1^{\rmL}
	\la{\bf Q}(0)\ra
},
\label{Q^m}
\end{eqnarray}
where
${\bf G}_\gamma(z)$,
${\bf R}_j^{\rmL}$, and
${\bf V}$ are $256\times 256$ matrices representing operators ${\cal G}_\gamma(z)$, and $R^{\rmL}_j$, ${\cal L}_{\rm int}$, respectively, with elements given by 
\begin{equation}
\fl
[{\bf G}_\gamma(z)]_{kl}
=
\Trace \left\{ Q_l^\dagger {\cal G}_\gamma Q_k \right\}
, \,
[{\bf R}_j^{\rmL}]_{kl}
=
\Trace \left\{ Q_l^\dagger R_j^{\rmL} Q_k \right\}
, \,
[{\bf V}]_{kl}
=
\Trace \left\{ Q_l^\dagger \Liou_{\rm int} Q_k \right\}
,
\nonumber
\end{equation}
while $	[{\bf Q}(0)]_k=\Trace \left\{ Q^\dagger_k|1\rangle_1\langle 1|_1\otimes |1\rangle_2\langle 1|_2\right\}$ is a vector with 256 elements.
Let us finally represent the Liouvillians $\Liou_{\rm int}^{(1)}$ and  $\Liou_{\rm int}^{(2)}$, see \eref{Lab1} and \eref{Lab2}, 
with the matrices
\begin{equation}
[{\bf V}^{(1)}]_{kl}=\Trace \left\{ Q_l^\dagger \Liou^{(1)}_{\rm int} Q_k \right\}, \quad
[{\bf V}^{(2)}]_{kl}=\Trace \left\{ Q_l^\dagger \Liou^{(2)}_{\rm int} Q_k \right\}. 
\label{V1V2}	
\end{equation}


\subsubsection{Configurational average.}
\label{sec:disorder}
The configuration or disorder average in \eref{Q^m} is necessary to capture the impact of changes of the random atomic
configuration during the fluorescence detection (section \ref{sec:separation}).
As a result of averaging, all terms carrying position-dependent phases
vanish. In addition to interference contributions with $\alpha\neq \beta$ in equation \eref{numer_I} which are already excluded in \eref{Q^m}, the disorder average obliterates all terms with  ${\bf r}_{\alpha 0}$--dependent phases---which are due to the action of the two incoming  laser pulses (see equations \eref{varphij}, \eref{phase_R1}). 
What remains are terms carrying position-independent phases that originate 
in single scattering processes, independent of ${\bf V}$, and double scattering processes, quadratic in ${\bf V}$. These single and double scattering contributions 
have weights $|g(\bar{\xi})|^0=1$ and
$|g(\bar{\xi})|^2=1/\bar{\xi}^2$, respectively. 

Furthermore, only particular products of the matrix elements of
$\Tensor{T}(\bar{\xi}, \uvec{n})$ (which enter \eref{Q^m} at second order in $\Liou_{\rm int}$, through \eref{L_ab})  contribute to the average signal upon
integration over the isotropic distribution of the vector $\uvec{n}$
\cite{shatokhin2006,ketterer2014}:
$
  \ConfAvg{
    \Tensor{T}_{kl}
    \Tensor{T}{}^*_{mn}
  }
  \neq
  0
$
if and only if the indices $k, l, m, n$ are pairwise identical, cf.~\ref{app:tensorT}. This amounts to the suppression, in the average fluorescence intensity, of all 
contributions stemming from the  interference of atomic excited state sublevels with different magnetic quantum numbers. 

Finally, note that through the matrix ${\bf R}^\rmL_j$ the atomic correlation functions carry phases $\Delta_\alpha t_j$ proportional to the Doppler shifts, by virtue of \eref{varphij}, \eref{phase_R1}. Unlike the uniform spatial distribution of the atoms, the Doppler shifts $\Delta_\alpha$ are non-uniformly Maxwell-Boltzmann distributed. Therefore, the result of averaging over $\Delta_\alpha$  is different from that of averaging over random atomic positions; it would not eliminate the terms including velocity-dependent phases, but instead result in inhomogeneous line broadening \cite{loudon} of MQC spectra \cite{ames2021}. In our present contribution, we  ignore this effect and set the Doppler shifts in equation \eref{varphij}  to zero, while focusing on the general structure and physical origin of the resonances of 1QC and 2QC spectra.

The final result after configuration averaging is a number of terms which contribute to the single and double scattering intensities. More precisely, these terms, derived from the vectors
$
\ConfAvg{ \la
	\tilde{\bf Q}^{[0]}(z_1,z_2,\varphi_1,\varphi_2)
	\ra }
$
and
$
\ConfAvg{ \la
	\tilde{\bf Q}^{[2]}(z_1,z_2,\varphi_1,\varphi_2)
	\ra }
$ in equation \eref{Q^m}, yield the two-dimensional Laplace images of the single and double scattering transient intensities denoted, respectively, by $\tilde{I}^{[0]}_{\uvec{k}}(z_1,z_2,\varphi_{1},\varphi_2)$ and $\tilde{I}^{[2]}_{\uvec{k}}(z_1,z_2,\varphi_{1},\varphi_2)$. 

We can draw general conclusions about the dependence of $\tilde{I}^{[0]}_{\uvec{k}}(z_1,z_2,\varphi_{1},\varphi_2)$ and $\tilde{I}^{[2]}_{\uvec{k}}(z_1,z_2,\varphi_{1},\varphi_2)$
on the phases $\varphi_j$. This is possible because, by \eref{varphij}, the $\varphi_j$ ($j=1,2$) are linear functions on the random initial positions ${\bf r}_{0\alpha}$ of the first or second atom, that is $\varphi_j=\varphi_{j\alpha}$ ($\alpha=1,2$). Consequently, the average functions $\tilde{I}^{[n]}_{\uvec{k}}(z_1,z_2,\varphi_{1},\varphi_2)$ ($n=0,2$) vanish unless they depend solely on a phase factor proportional to the difference $\varphi_{21}=\varphi_{2\alpha}-\varphi_{1\alpha}=\omega_0\tau+w_{21}\tau_m$ or, trivially, are independent of $\varphi_{j\alpha}$. 
More explicitly, the non-vanishing intensities $\tilde{I}^{[n]}_{\uvec{k}}(z_1,z_2,\varphi_{1},\varphi_2)
=
\tilde{I}^{[n]}_{\uvec{k}}(z_1,z_2,\varphi_{21})
$ can then be written as
\begin{eqnarray}
\tilde{I}^{[0]}_{\uvec{k}}(z_1,z_2,\varphi_{21})
&=&
\sum_{l=-1}^1 \tilde{I}^{[0]}_{l,\uvec{k}}(z_1,z_2)\rme^{\rmi \varphi_{21} l},
\label{I^0_expansion}\\
\tilde{I}^{[2]}_{\uvec{k}}(z_1,z_2,\varphi_{21})
&=&
\sum_{l=-2}^2 \tilde{I}^{[2]}_{l,\uvec{k}}(z_1,z_2)\rme^{ \rmi \varphi_{21} l}.
\label{I^2_expansion}
\end{eqnarray}
The summation range in \eref{I^0_expansion} is a consequence of equation \eref{Rj-action-ground-state}:
since, for $n=0$, the time evolution \eref{Q^m} is local in the Hilbert space of each atom,
 individual atomic operators can depend only on the phase tags $\rme^{\pm \rmi \varphi_{21}}$, and not on $\rme^{\pm 2 \rmi \varphi_{21}}$. 
In contrast, fluorescence intensity emerging due to double scattering can carry phases $\pm 2\varphi_{21}$ in  $\tilde{I}^{[2]}_{\uvec{k}}(z_1,z_2,\varphi_{21})$, giving rise to 2QC signals.

Taking the inverse Laplace transforms (see \eref{inverseL}) of the functions defined by  \eref{I^0_expansion} and \eref{I^2_expansion}
with respect to $z_1$ and $z_2$, we obtain the associated time-domain expressions $I^{[0]}_{l,\uvec{k}}(\tau,t_{\rm fl},\varphi_{21})$ and $I^{[2]}_{l,\uvec{k}}(\tau,t_{\rm fl},\varphi_{21})$, 
\begin{eqnarray}
	I^{[0]}_{\uvec{k}}(\tau,t_{\rm fl},\varphi_{21})&=&\sum_{l=-1}^1 I^{[0]}_{l,\uvec{k}}(\tau,t_{\rm fl})\rme^{\rmi \varphi_{21} l},\label{It^0_expansion}\\
	I^{[2]}_{\uvec{k}}(\tau,t_{\rm fl},\varphi_{21})&=&\sum_{l=-2}^2 I^{[2]}_{l,\uvec{k}}(\tau,t_{\rm fl})\rme^{\rmi \varphi_{21} l},
	\label{It^2_expansion}
\end{eqnarray}
where $I^{[0]}_{l,\uvec{k}}(\tau,t_{\rm fl})$ and $I^{[2]}_{l,\uvec{k}}(\tau,t_{\rm fl})$ are exponentially decaying functions of $\tau$ and $t_{\rm fl}$  (recall our discussion in \sref{sec:nonint_multi}), related to $\tilde{I}^{[n]}_{l,\uvec{k}}(z_1,z_2)$  by the two-dimensional Laplace transform:
\begin{equation}
\tilde{I}^{[n]}_{l,\uvec{k}}(z_1,z_2)=\int_0^\infty \dif{\tau}\rme^{-z_1\tau}\int_0^\infty
\dif{t_{\rm fl}} \rme^{-z_2 t_{\rm fl}} I^{[n]}_{l,\uvec{k}}(\tau,t_{\rm fl}).
\label{Iz1z2_It1t2}
\end{equation}

The total transient intensity is the sum of \eref{It^0_expansion} and \eref{It^2_expansion},
\begin{equation}
I_{\uvec{k}}(\tau,t_{\rm fl},\varphi_{21})
=
I^{[0]}_{\uvec{k}}(\tau,t_{\rm fl},\varphi_{21})
+
I^{[2]}_{\uvec{k}}(\tau,t_{\rm fl},\varphi_{21})
,
\label{I_k}
\label{I_trans}
\end{equation}
which can be used to deduce 1QC and 2QC spectra, as outlined in section \ref{sec:spectroscopy}, and discussed below.

\subsubsection{Signal demodulation.}
\label{sec:demod}
The goal of the demodulation procedure is to extract only those specific components
from the fluorescence signal, which oscillate at defined integer multiples $\kappa$ of the modulation frequency $w_{21}$.  
The spectra of these signals, generally referred to as multiple quantum coherences  (MQC), are 
extracted from \eref{intensity}, after replacement of $\phi_1$ and $\phi_2$ by the phase difference $\varphi_{21}$, via the following two-fold, half-sided 
Fourier transforms,
\begin{eqnarray}
S_{\uvec{k}}(\omega; \kappa)
&=&
\lim_{\FLI \to 0}
\frac{\FLI}{\sqrt{2\pi}}
\int_0^\infty \dif{\tau}_m
\rme^{-(\FLI+\rmi \kappa w_{21})\tau_m}
\int_0^\infty \dif{\tau}
\rme^{-\rmi \omega \tau}
\nonumber \\
&&
\times
\int_{0}^\infty \dif{t_{\rm fl}}
I_{\uvec{k}}(\tau, t_{\rm fl}, \varphi_{21}), 
\label{dem_signal_explicit1}
\end{eqnarray}
with $\kappa=1, 2$ the MQC orders here considered.
\eref{I_trans}, with $I^{[0]}_{\uvec{k}}(\tau,t_\mathrm{fl},\varphi_{21})$ and $I^{[2]}_{\uvec{k}}(\tau,t_\mathrm{fl},\varphi_{21})$ 
from \eref{It^0_expansion} and \eref{It^2_expansion}, into \eref{dem_signal_explicit1}, and integration over $\tau_m$ yields the result:
\begin{eqnarray}
S_{\uvec{k}}(\omega; \kappa)
&=&
\lim_{\FLI \to 0}
\frac{\FLI}{\sqrt{2\pi}}\left[\sum_{l=-1}^1
\int_0^\infty\dif{\tau} \rme^{-\rmi(\omega-l \omega_0)\tau}
\int_{0}^\infty\dif{t_\mathrm{fl}}
\frac{I^{[0]}_{l,\uvec{k}}(\tau,t_\mathrm{fl})}{\FLI+
\rmi(\kappa-l)w_{21}} \right.\nonumber\\
&&\left.+
\sum_{l=-2}^2\int_0^\infty\dif{\tau} \rme^{-\rmi(\omega-l \omega_0)\tau}
\int_{0}^\infty\dif{t_\mathrm{fl}}\frac{I^{[2]}_{l,\uvec{k}}(\tau,t_\mathrm{fl})}{\FLI+\rmi(\kappa-l)w_{21}}
\right].
\label{S_interm}
\end{eqnarray}
The limit ${\cal F}\to 0$ selects only terms corresponding to $l=\kappa$ in \eref{S_interm}, and is equivalent to \eref{Iz1z2_It1t2} evaluated at $z_1=\rmi(\omega-\kappa\omega_0)$ and $z_2=0$. We thus arrive at the final expressions  for the 1QC and 2QC spectra, which exhibit resonances at $\omega=\omega_0$ and $\omega=2\omega_0$, respectively:
\begin{eqnarray}
	S_{\uvec{k}}(\omega; 1)
	&=&
	\frac{1}{\sqrt{2\pi}}\tilde{I}^{[0]}_{1,\uvec{k}}(\rmi \bigl(\omega-\omega_0),0\bigr)+	\frac{1}{\sqrt{2\pi}}\tilde{I}^{[2]}_{1,\uvec{k}}\bigl(\rmi (\omega-\omega_0),0\bigr),\label{S1}\\
	S_{\uvec{k}}(\omega; 2)
	&=&
		\frac{1}{\sqrt{2\pi}}\tilde{I}^{[2]}_{2,\uvec{k}}\bigl(\rmi (\omega-2\omega_0),0\bigr).\label{S2}
\end{eqnarray}

\begin{figure*}
	\center
	\includegraphics[width=0.8\textwidth]{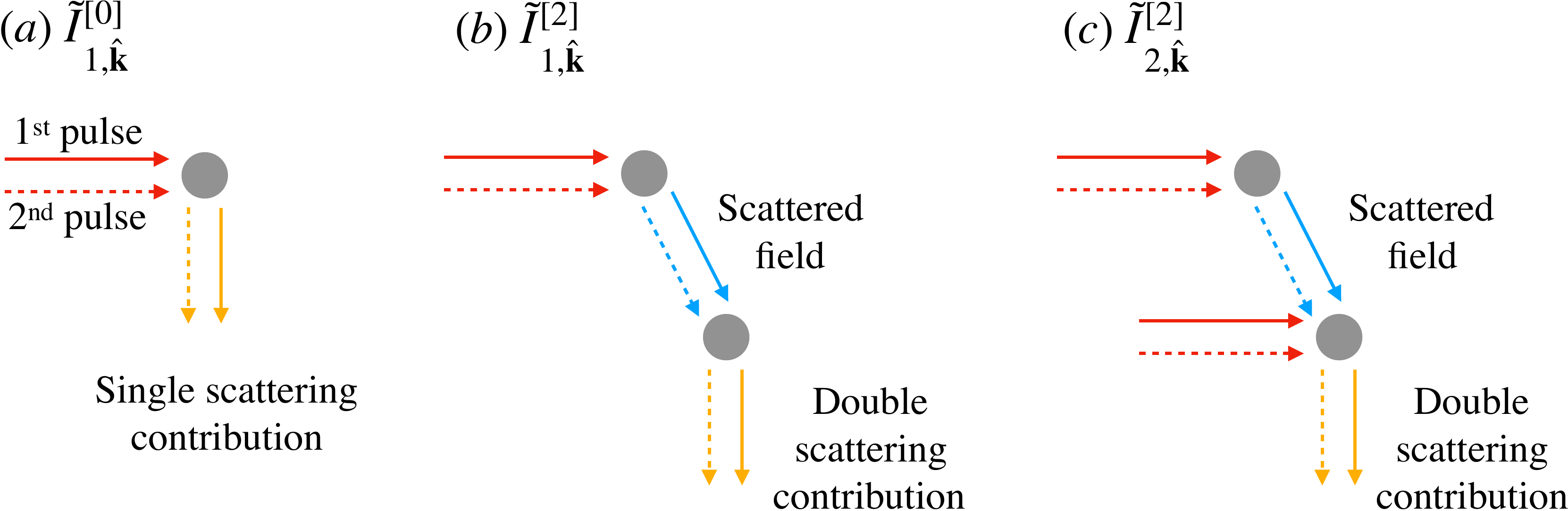}
	\caption{
		Diagrammatic representation of single (a) and double (b,c) scattering processes contributing to intensities $\tilde{I}^{[0]}_{1,\uvec{k}}$, $\tilde{I}^{[2]}_{1,\uvec{k}}$, and $\tilde{I}^{[2]}_{2,\uvec{k}}$, respectively (see \eref{S1}, \eref{S2}). Horizontal red lines indicate the incoming laser pulses, vertical yellow lines the emitted fluorescence, oblique blue lines the scattered fields via dipole-dipole interactions. 
		Solid and dashed lines 
		indicate single field amplitudes and their complex conjugates (see also \fref{fig:setup}). The difference between the double scattering processes (b) and (c) is that, in the former case,  
		fluorescence stems from the excited state sublevels of the second atom that are driven by the scattered, but 
	\emph{not} by the incident field,
		while, in the latter case,   
		by \emph{both} the scattered and the incident field. As discussed in \sref{sec:spectra}, processes (a,b) and (c) are associated with 1QC and 2QC signals, respectively. 				
	}
	\label{fig:scattering}
\end{figure*}

To summarise, through \eref{S1} and \eref{S2} we connect 1QC and 2QC spectra as measured by methods of phase-modulated spectroscopy to specific components of the fluorescence intensity emitted by two atoms. The intensity is expressed via \eref{numer_I} (with $N=2$ and, as a consequence of the configuration average over the atomic random positions, $\alpha=\beta$) through the excited state populations of {\it single} atoms that may be independent or interacting with each other by exchanging a single photon. 
Furthermore, while equation \eref{S1} indicates that the 1QC signal is fed by both, single and double scattering processes,  
equation \eref{S2} shows that 2QC signal can stem solely from double scattering, see \fref{fig:scattering}. In other words, whereas 1QC signals may be observed on independent particles, 2QC signals are always an unambiguous proof of interatomic interactions. Thus the origin of 2QC signals in phase-modulated spectroscopy on disordered samples \cite{lukas_bruder15,Bruder_2019} is substantially different from the one outlined in \cite{Mukamel_2016}, where 2QC signals stem from two-atom observables of non-interacting atoms.  

The dipolar interactions are represented by excitation exchange and collective decay processes encapsulated in the Liouvillians 
${\cal L}_{\alpha\beta}^{(1)}$ and ${\cal L}_{\alpha\beta}^{(2)}$ defined by \eref{Lab1} and \eref{Lab2}. In \sref{sec:spectra}, we present the spectra \eref{S1} and \eref{S2} 
and elucidate the significance of these Liouvillians for the emergence of 1QC and 2QC signals, which explains a number of experimentally observed features of MQC 
spectra \cite{Bruder_2019}.   
%

\section{Spectra of single and double quantum coherence signals}
\label{sec:spectra}
\begin{figure*}
	\includegraphics[width=\textwidth]{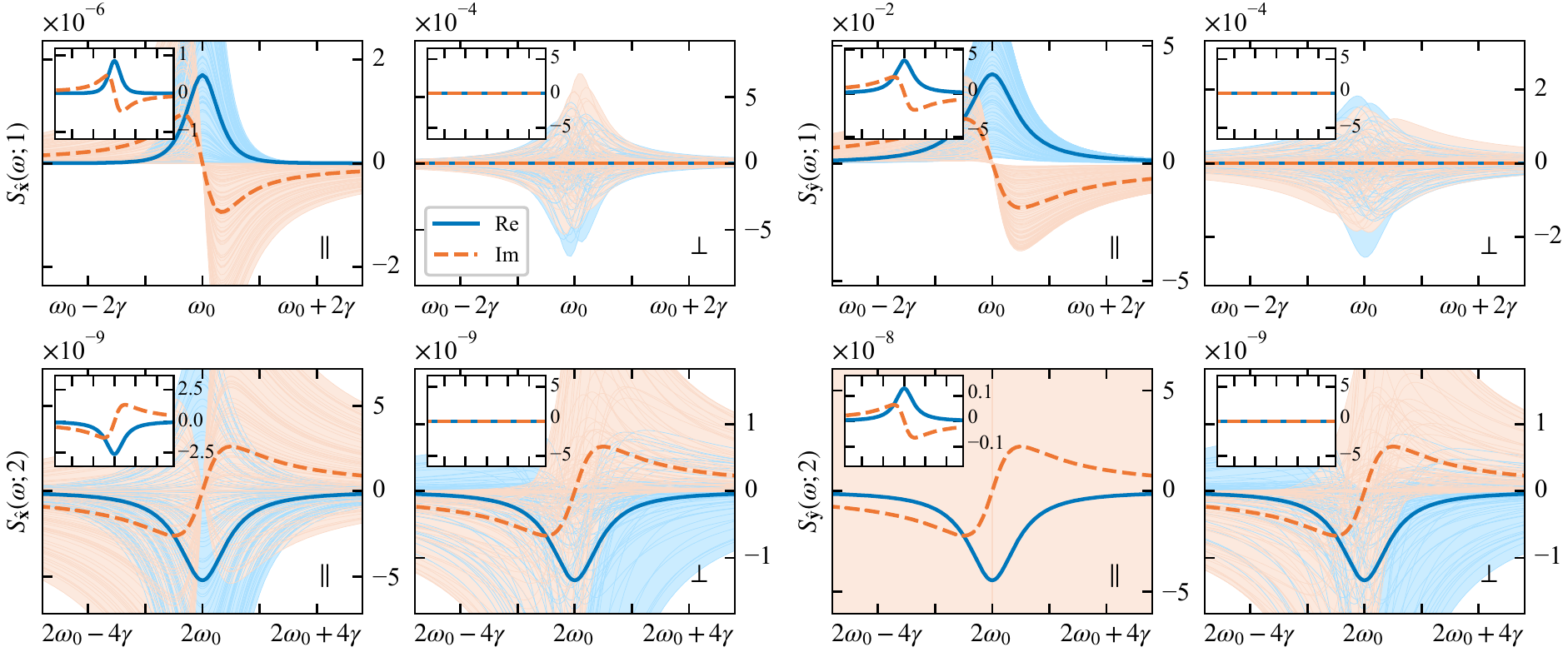}
	\caption{Real (solid lines) and imaginary (dashed lines) parts of the
		MQC spectra $S_{\uvec{k}} (\omega;\kappa)$
		[in units of spectral density integrated over the fluorescence detection time, $f^2/\gamma^2$; see \eref{numer_I} and \eref{dem_signal_explicit1}],
		for detection directions $\uvec{k}=\uvec{x}$ (left) and
		$\uvec{k}=\uvec{y}$ (right),
		and for quantum coherences of order $\kappa=1$ (top) and $\kappa=2$ (bottom).
		Symbols $\parallel$ and $\perp$ indicate
		parallel ($\uvec{x}$-$\uvec{x}$)
		and perpendicular ($\uvec{x}$-$\uvec{y}$)
		pump-probe polarizations, respectively. 
		Parameter values match the experimental ones \cite{Bruder_2019}:
		 average
		interatomic distance $\bar{r}\approx \SI{10}{\micro\meter}$ (which
		corresponds to a particle density $\rho =\SI{e8}{\per\cm\cubed}$),
		laser pulse areas $\vartheta=\num{0.14}\pi$ and durations
		$\sigma= \SI{21}{fs}$ (tuned to resonance with the D2-line
		of $^{87}$Rb atoms \cite{steckRubidium87Line2015}), transition wavelength $\lambda_0=\SI{790}{nm}$,
		and spontaneous decay rate
		$\gamma \approx 2\pi \times \SI{6,067}{MHz}$.
		This choice of parameters results in a mean scaled interatomic distance
		$\bar{\xi}=k_0\bar{r}\approx 80$, with $k_0$ the atomic transition's wave number.
	 Thin lines in the background show MQC spectra (with the same color coding for the real and imaginary parts as the average spectra) for 100 random configurations $(\xi, \hat{\bf n})$, with $\xi$ distributed uniformly in the interval $
	67.2
		<
	\xi
		<
		92.8
		$, and $\hat{\bf n}$ distributed isotropically. The insets show the signals obtained by substituting $\Tensor{T} \mapsto \rmi \Tensor{\Omega}$, which amounts to ignoring the collective decay processes.
 	}
	\label{fig:spectra}
\end{figure*}

In \fref{fig:spectra} we plot the single and double coherence spectra $S_{\uvec{k}}(\omega;1)$ and $S_{\uvec{k}}(\omega;2)$, given by \eref{S1} and \eref{S2}, for
$\uvec{k}=\uvec{x}, \uvec{y}$, and for parallel ($\uvec{x}$-$\uvec{x}$) and perpendicular ($\uvec{x}$-$\uvec{y}$) pump-probe polarization channels (in brief, $\parallel$ and $\perp$ channels). 
As discussed in section \ref{sec:disorder}, these signals emerge from the configuration average.
For illustration, the background of
each panel in \fref{fig:spectra} shows MQC spectra stemming from
pairs of atoms at fixed locations. These were calculated
by omitting
the average~$\ConfAvg{\ldots}$ in \eref{Q^m} and, instead, by considering randomly drawn 
configurations~$(\xi, \uvec{n})$, which determine the tensor
$\Tensor{T}$ and the phase factor $\exp(\rmi {\bf k}\cdot{\bf r}_{12})$---as detailed in the caption of \fref{fig:spectra}.
The deviations between background and averaged spectra are striking for both parallel and
perpendicular pulses:  the
single realizations have little resemblance with the true (configuration averaged) line shapes, from which they oftentimes differ by orders of magnitude.
The absence of the configuration average constitutes one of several reasons for a substantial qualitative and quantitative mismatch between the model including two fixed molecules \cite{Li_2017} and the observations reported in \cite{Bruder_2019}. Further reasons will be identified below. Henceforth, we will discuss our configuration-averaged spectra. 

We stress that, while our general analytical solutions, derived from \eref{Q2} and \eref{Q^m}, assume that dipolar interactions act during both, the interpulse delay and fluorescence harvesting, we have checked numerically that the 1QC and 2QC spectra can be fully understood if we include the interatomic interactions only during the fluorescence detection stage. This is reasonable, given the typically short duration of $\sim \SI{10}{\pico\second}$ of the interpulse delay and the weakness of the dipolar coupling. Therefore, throughout this section we interpret the emergence of 2QC spectra (as much as contributions to 1QC spectra) as the result of dipole-dipole interactions during the long time scale, i.e.\ \emph{after} the atoms have interacted with the two incoming laser pulses.      

In the  $\parallel$ channel, the real and imaginary parts of the 1QC spectra exhibit, respectively, absorptive and dispersive resonances at $\omega=\omega_0$.  In the here realized regime of relatively weak (perturbative) driving fields, these line shapes are consistently proportional to the real and imaginary parts of the linear susceptibility~\cite{Tekavec2006}.
For both choices of pump-probe polarizations and observation directions, the 2QC spectra feature qualitatively identical line shapes centred at $\omega=2\omega_0$, albeit with opposite sign to the 1QC spectra. The sign flip of the 2QC spectra can be understood by recalling that they originate from double scattering. Thereby, each atom effectively interacts with four fields: two laser pulses and two far-fields scattered by the other atom, see \fref{fig:scattering} (c). This nonlinear four-wave mixing process is described by the third-order susceptibility \cite{mukamel_book,boyd_book}, whose sign is opposite to that of the linear susceptibility \cite{boyd_book}. We would like to stress that the sign flip of the 2QC spectra was experimentally observed \cite{Bruder_phd}, but its interpretation has been lacking. 

In the $\parallel$ channel, 
the peak magnitude of the 1QC signal for $\uvec{k}=\uvec{x}$ is four orders of magnitude smaller
than for $\uvec{k}=\uvec{y}$, see \fref{fig:spectra}. This significant difference arises because the signal for $\uvec{k}=\uvec{y}$ is dominated by single scattering from individual atoms, as illustrated by \fref{fig:scattering} (a) (i.e., by the $\tilde{I}^{[0]}_{1,\uvec{k}}$ component in \eref{S1}). This contribution to the single coherence signal is independent of the interatomic distance. 
For $\uvec{k}=\uvec{x}$, on the contrary, the atoms in their $|\rmx\rangle$ state cannot directly emit in the $\hat{\bf x}$-direction. Hence, the signal is due to double
scattering contribution $\tilde{I}^{[2]}_{1,\uvec{k}}$ (see \fref{fig:scattering} (b)), wherein the excitation is transferred from the state $|\rmx\rangle$ of one atom to the state $|\rmy\rangle$ or $|\rmz\rangle$ of the other --- which was in its ground state prior to the interaction. Such an interaction process, involving the exchange of an excitation between the atoms, is mediated solely by the Liouvillian ${\cal L}_{\alpha\beta}^{(1)}$, see \eref{Lab1}.\footnote{Henceforth, while discussing pathways contributing to 2QC signals, we refer to the atomic variables extracted from the solution of the matrix equation \eref{Q^m} rather than the operator equation \eref{Q2}.  Therefore, instead of ${\cal L}_{\alpha\beta}^{(1)}$ and ${\cal L}_{\alpha\beta}^{(2)}$, we resort to their associated matrices ${\bf V}^{(1)}$ and ${\bf V}^{(2)}$, see \eref{V1V2}.} The total average probability of this process is proportional to $ \ConfAvg{\Tensor{T}_{xr}\Tensor{T}{}^*_{xr}}=\ConfAvg{\Tensor{\Omega}_{xr}\Tensor{\Omega}_{xr}}+\ConfAvg{\Tensor{\Gamma}_{xr}\Tensor{\Gamma}_{xr}}$ ($r=y,z$), which yields a nonzero contribution $\sim 1/\bar{\xi}^2=0.00016$ for $\bar{\xi}=80$ (see caption of figure \ref{fig:spectra}), justifying the difference, by four orders of magnitude, between the 1QC signals in the $\uvec{x}$ and $\uvec{y}$ directions. 

Note the complete absence in our data of the 1QC signal in the $\perp$ channel
(see 2$^{\rm nd}$ and 4$^{\rm th}$ panel in the top row of \fref{fig:spectra}), corroborating the experiment \cite{Bruder_2019}.
This is due to orthogonally polarized laser pulses imprinting the
correct phase for demodulation only on the Zeeman coherences (see  \sref{sec:ultrashort-pulses}), i.e.\ the off-diagonal elements $\langle \sigma_{xy}^\alpha\rangle$ in the excited state manifold of the atomic
density matrix, which do not directly
fluoresce.
While the dipole-dipole interaction can transfer the Zeeman coherence of atom $\alpha$ to the excited state population $\langle \sigma_{rr}^\beta\rangle$ ($r \in \{ x, y, z \}$) of the initially unexcited atom $\beta$ via the matrix ${\bf V}^{(1)}$ (see \sref{sec:asymp_dd} and equation \eref{V1V2}), 
the probability of this process is proportional to
$\Tensor{T}_{ x r}\Tensor{T}{}^*_{ y r}$,
which vanishes upon disorder averaging
(see \sref{sec:disorder} and \ref{app:tensorT}). It must be emphasized that an appropriate assessment of MQC signals in the $\perp$ channel is only possible when the vector character of light fields and dipoles is accounted for, and is therefore out of reach for scalar atom models ~\cite{Dai_2012,Mukamel_2016,Li_2017,Gao:16,PhysRevLett.120.233401,Yu:19a}.

\begin{figure*}
	\includegraphics[width=\textwidth]{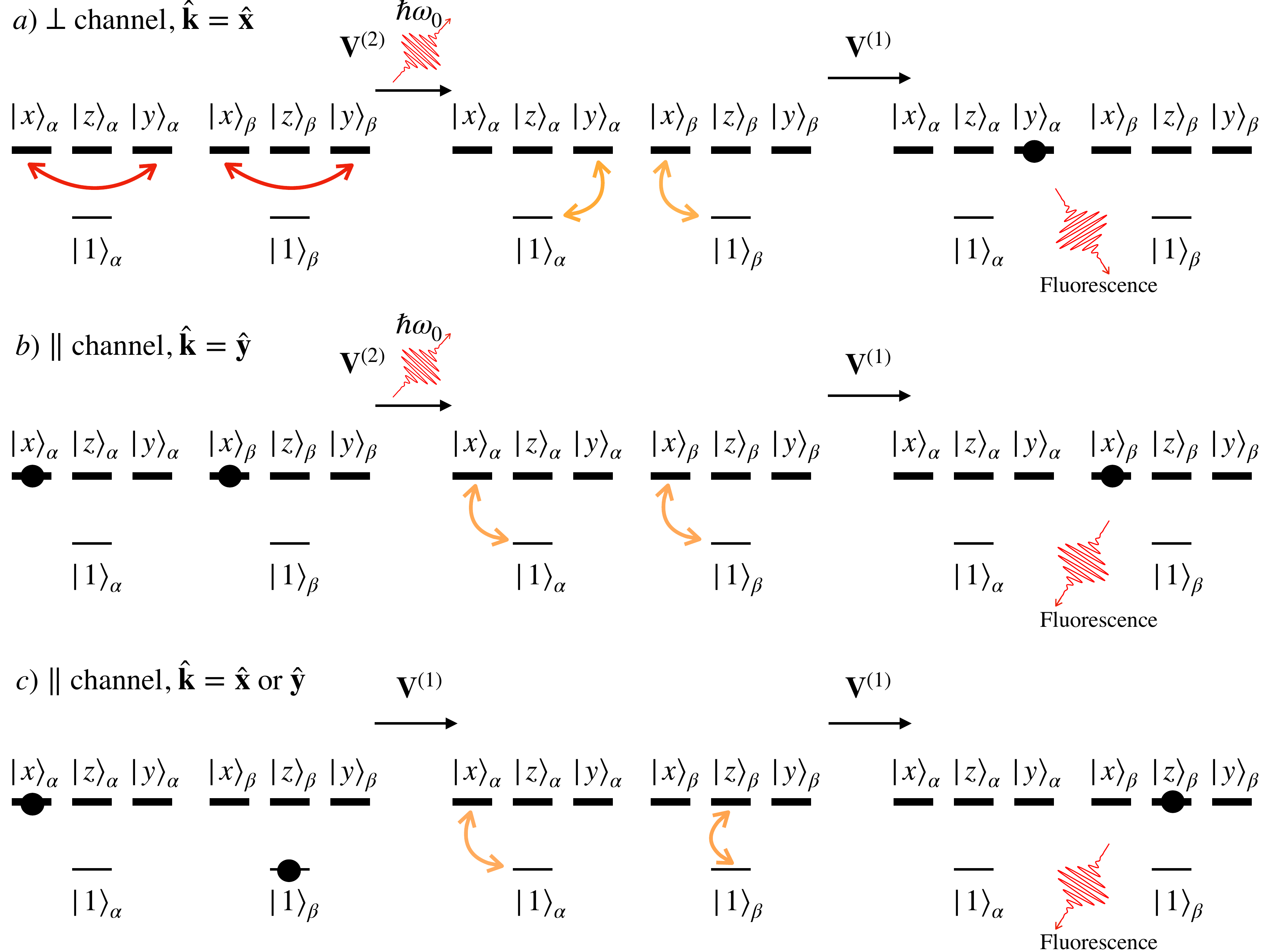}
	\caption{Excitation pathways describing the transformation of the initial 
		two-atom variables with the correct phase for demodulation of 2QC signals (see \sref{sec:ultrashort-pulses} and  \fref{fig:coherences}) to the excited state populations 
		of the atoms which give rise to fluorescence signals. Only those variables that contribute to 2QC signals are indicated. (a) In the $\perp$ channel and for the 
		observation direction $\uvec{k}=\uvec{x}$, Zeeman coherences $\langle \sigma_{xy}^\alpha \rangle\langle\sigma_{xy}^\beta\rangle$ (red arcs) of independent atoms  
		are transformed, by the collective decay process 
		mediated by the matrix ${\bf V}^{(2)}$, and accompanied by the emission of a single photon, into electronic coherences (orange arcs), e.g., 
		$\langle \sigma^\alpha_{1y}\sigma^\beta_{x1}\rangle$. By the interaction represented by 
		the matrix ${\bf V}^{(1)}$, the atomic state $|\rmy\rangle_\alpha$ is excited, while atom $\beta$ is de-excited, which is described by the correlator 
		$\langle \sigma^\alpha_{yy}\sigma^\beta_{11}\rangle$ (black dot). The fluorescence emitted by atom $\alpha$ contributes to the 2QC signal. 
		(b) In the $\parallel$ channel and for the observation direction $\uvec{k}=\uvec{y}$, the pathway indicates the following sequence of processes: 
		$\langle \sigma^\alpha_{xx}\rangle\langle\sigma^\beta_{xx}\rangle\overset{{\bf V}^{(2)}}\longrightarrow \langle\sigma^\alpha_{1x}\sigma^\beta_{x1}\rangle\overset{{\bf V}^{(1)}}\longrightarrow \langle\sigma^\alpha_{11}\sigma^\beta_{xx}\rangle$.
		(c) In the $\parallel$ channel and for the observation directions $\uvec{k}=\uvec{y}$ or $\uvec{x}$, additional pathways are mediated solely by the matrices 
		${\bf V}^{(1)}$: $\langle \sigma^\alpha_{xx}\rangle\langle\sigma^\beta_{11}\rangle\overset{{\bf V}^{(1)}}\longrightarrow \langle\sigma^\alpha_{1x}\sigma^\beta_{z1}\rangle\overset{{\bf V}^{(1)}}\longrightarrow \langle\sigma^\alpha_{11}\sigma^\beta_{zz}\rangle$.
	}
	\label{fig:pathways}
\end{figure*}
In the same ($\perp$) pump-probe configuration, 2QC signals are however present -- in qualitative agreement with experiment \cite{Bruder_2019}.
They originate from the Zeeman coherences
$\langle \sigma_{xy}^\alpha \sigma_{xy}^\beta\rangle=\langle \sigma_{xy}^\alpha \rangle\langle\sigma_{xy}^\beta\rangle$ (as discussed above, this two-atom correlation function factorizes because the atoms have not yet interacted with each other)
induced by the incoming pulses in both atoms (recall our discussion in \sref{sec:ultrashort-pulses} accompanying \fref{fig:coherences}(b)).
Given such a doubly-excited state, the matrix ${\bf V}^{(2)}$ (see equation \eref{V1V2})
mediates a collective decay process (see \fref{fig:pathways} (a)), where one photon is emitted by
both atoms, transforming the initial two-atom product state with Zeeman coherences into a 
mixed (possibly entangled) state exhibiting electronic coherences of the atoms, e.g.\ $\langle \sigma^\alpha_{1y}\sigma^\beta_{x1}\rangle$. 
This process is linear in
$
  \Tensor{\Gamma}_{x y}
$
and, hence, does not directly contribute to the average 2QC signal.
However, the electronic coherence of one atom can be further transformed via the matrix ${\bf V}^{(1)}$ into
the excited state population of the other atom, such as $\langle \sigma^\alpha_{11}\sigma^\beta_{xx}\rangle$ or $\langle \sigma^\alpha_{yy}\sigma^\beta_{11}\rangle$ (see \fref{fig:pathways} (a)).
Although ${\bf V}^{(1)}$ is linear in both $\Tensor{\Omega}$ and $\Tensor{\Gamma}$, the total average probability of preparing such a state is proportional
to
$
  \ConfAvg{
    \Tensor{\Gamma}_{y x}
    \Tensor{\Gamma}_{x y}
  }
  \neq 0
$ (while $
\ConfAvg{
	\Tensor{\Gamma}_{y x}
	\Tensor{\Omega}_{x y}
}
= 0$, cf.~\ref{app:tensorT}). 
Subsequent fluorescence from such a state, consisting of one atom in the excited and the other one in the ground state, \emph{does} give rise to a non-vanishing average
2QC signal with the same intensity in directions $\hat{\bf y}$ and $\hat{\bf x}$, respectively.  

In the $\parallel$ channel, the correct phase for demodulation of 2QC signals is carried by combinations of the two-atom correlators $\langle \sigma_{xx}^\alpha \sigma_{xx}^\beta\rangle=\langle \sigma_{xx}^\alpha\rangle \langle \sigma_{xx}^\beta\rangle$, $\langle \sigma_{11}^\alpha \sigma_{xx}^\beta\rangle=\langle \sigma_{11}^\alpha\rangle\langle \sigma_{xx}^\beta\rangle$, and $\langle \sigma_{xx}^\alpha \sigma_{11}^\beta\rangle=\langle \sigma_{xx}^\alpha\rangle\langle \sigma_{11}^\beta\rangle$ (see section \ref{sec:ultrashort-pulses}). A possible pathway that can lead to the contribution of the correlator 
$\langle \sigma_{xx}^\alpha \sigma_{xx}^\beta\rangle$ to the fluorescence signal in direction $\uvec{y}$ is the following: By means of two interactions generated by  the matrices ${\bf V}^{(2)}$ and  ${\bf V}^{(1)}$ (see \fref{fig:pathways} (b)), one atom is prepared in the ground and the other in the excited state $|x\rangle$. This is analogous to how the Zeeman coherences $\langle \sigma_{xy}^\alpha \sigma_{xy}^\beta\rangle$ can contribute to the signal in the $\perp$ channel. The total average probability of preparing such state is $\propto
\ConfAvg{
	\Tensor{\Gamma}_{x x}
	\Tensor{\Gamma}_{x x}
}
$, and a 2QC signal due to the correlator $\langle \sigma_{11}^\alpha \sigma_{xx}^\beta\rangle$ can be observed in direction $\hat{\bf y}$. The correlation functions $\langle \sigma_{11}^\alpha \sigma_{xx}^\beta\rangle$ and $\langle \sigma_{xx}^\alpha \sigma_{11}^\beta\rangle$, instead, can contribute to 2QC spectra  via two interactions generated by the matrix ${\bf V}^{(1)}$ (for the correlation function $\langle \sigma_{xx}^\alpha\rangle \langle\sigma_{11}^\beta\rangle$, a possible pathway is displayed in \fref{fig:pathways} (c)). Excited state populations are thereby transferred from atom $\alpha$ to atom $\beta$ or vice versa (\fref{fig:pathways} (c) shows a pathway leading to the correlator $\langle \sigma_{11}^\alpha \sigma_{zz}^\beta\rangle$). The total average probability of such processes is $\propto \ConfAvg{
	\Tensor{\Gamma}_{x r}
	\Tensor{\Gamma}_{x r}
}+
\ConfAvg{
	\Tensor{\Omega}_{x r}
	\Tensor{\Omega}_{x r}
}
$ ($r \in \{x, y, z\}$), and the signal can be observed both in directions $\hat{\bf x}$ (for $r=y,z$) and $\hat{\bf y}$ (for $r=x,z$). Thus, in the $\parallel$ channel, the intensity of 2QC signal includes the additional contribution of the collective decay processes in the $\hat{\bf y}$-direction of observation, which explains the difference in the magnitudes of the $S_{\uvec{x},\parallel}(\omega,2)$ and $S_{\uvec{y},\parallel}(\omega,2)$ spectra (see \fref{fig:spectra}). It is worth mentioning that the here described far-field dipole-dipole interaction processes mediated by real photons are somewhat akin to the interatomic Coulombic decay processes in molecular donor-acceptor systems \cite{Cederbaum1997,Hemmerich:2018wd}, where a donor undergoes an internal transition emitting a photon, which then ionizes an acceptor.   

To highlight the role of the collective decay processes even more, in the insets of \fref{fig:spectra} we show the analytical spectra calculated with the real part, $\Tensor{\Gamma}$, of the interaction tensor set to zero (i.e., modeling the radiative dipole-dipole interaction as purely Hamiltonian, $\propto \Tensor{\Omega}$, giving rise to collective level shifts, recall our discussion in \sref{sec:asymp_dd}). This setting retains only the interaction matrix ${\bf V}^{(1)}$, while the matrix ${\bf V}^{(2)}$ vanishes, see equations \eref{Lab1}, \eref{Lab2}, \eref{V1V2}, and so do the 2QC signals for perpendicularly polarized pulses. Furthermore, the magnitudes of the spectra $S_{\hat{\bf x}}(\omega;1)$ and $S_{\hat{\bf x}}(\omega;2)$ are reduced by a factor of two, in accordance with the facts that these signals are mediated solely by ${\bf V}^{(1)}$, hence, 
$\propto \ConfAvg{
	\Tensor{\Gamma}_{x r}
	\Tensor{\Gamma}_{x r}
}+
\ConfAvg{
	\Tensor{\Omega}_{x r}
	\Tensor{\Omega}_{x r}
}
$ (see the above paragraph),
and that $\ConfAvg{
	\Tensor{\Gamma}_{x r}
	\Tensor{\Gamma}_{x r}
}=
\ConfAvg{
	\Tensor{\Omega}_{x r}
	\Tensor{\Omega}_{x r}
}  
$ ($r=y,z$). 
As for the 2QC spectrum $S_{\hat{\bf y}}(\omega;2)$, its magnitude is diminished and its sign is flipped when the collective decay and excitation exchange processes proportional to the tensor $\Tensor{\Gamma}$ are switched off. Thus, using the static form of the dipole-dipole interaction~\cite{Dai_2012,Mukamel_2016,Li_2017,Gao:16,PhysRevLett.120.233401} (that is, completely ignoring $\Tensor{\Gamma}$) results in overlooking quantitative and/or qualitative features of 1QC and 2QC signals which were measured experimentally \cite{Bruder_2019,Bruder_phd}, in both $\perp$ and $\parallel$ channels, and for $\hat{\bf x}$ and $\hat{\bf y}$ observation directions. 

\begin{table}[tb]
	\centering
	\caption{%
		Peak amplitudes $\Re \{S_{\uvec{k}}(\kappa \omega_0;\kappa)\}$ (up to a common prefactor),
		at leading order in $\vartheta$ and  $1/\bar{\xi}$, for $\uvec{k}=\uvec{x}, \uvec{y}$, at demodulation order $\kappa$, and for
		pump-probe polarizations $\parallel$ and $\perp$  (see
		\fref{fig:spectra}).
	}
	\belowrulesep=1ex  
	\aboverulesep=1ex  
	\begin{tabular}{%
			>{$}c<{$}%
			>{$ \displaystyle \bgroup}c<{\egroup \displaystyle$}%
			>{$ \displaystyle \bgroup}c<{\egroup \displaystyle$}%
			>{$ \displaystyle \bgroup}c<{\egroup \displaystyle$}%
			>{$ \displaystyle \bgroup}c<{\egroup \displaystyle$}%
		}
		\toprule
		& \multicolumn{2}{c}{$\uvec{k}= \uvec{x}$}
		& \multicolumn{2}{c}{$\uvec{k} = \uvec{y}$}
		\\
		\cmidrule(lr){2-3}
		\cmidrule(lr){4-5}
		\kappa
		& \parallel & \perp
		& \parallel & \perp
		\\
		\midrule
		1
		&
		\frac{
			3
		}{
			10 \bar{\xi}^2
		}
		\vartheta ^2
		&
		0
		&
		\vartheta ^2
		&
		0
		\\[2ex]
		2
		&
		-
		\frac{
			3
		}{
			320 \bar{\xi}^2
		}
		\vartheta ^4
		&
		-
		\frac{
			3
		}{
			1280 \bar{\xi}^2
		}
		\vartheta ^4
		&
		-
		\frac{
			51
		}{
			640 \bar{\xi}^2
		}
		\vartheta ^4
		&
		-
		\frac{
			3
		}{
			1280 \bar{\xi}^2
		}
		\vartheta ^4
		\\
		\bottomrule
	\end{tabular}
	\label{tab1}
\end{table}

Finally, the origin of the different peak values of 1QC and 2QC signals lies in
their distinct scaling  with the pulse area $\vartheta$:
We find that $S_{\uvec{k}}(\omega;1)\propto \vartheta^2$, while
$S_{\uvec{k}}(\omega;2)\propto \vartheta^4$, see \tref{tab1} and \fref{fig:scattering}.
Different numerical prefactors for different polarization/observation channels
stem from the excitation pulses.
It is notable that, for parallel pump-probe polarization, both, the $S_{\uvec{x}, \parallel}(\omega_0;1)$ and $S_{\uvec{x}, \parallel}(2\omega_0;2)$ spectra, originate from double scattering and, consequently, scale as $1/\bar{\xi}^2$. Therefore, the ratio
$
  \Re \{S_{\uvec{x}, \parallel}(2\omega_0;2)\}
  /
  \Re \{S_{\uvec{x}, \parallel}(\omega_0;1)\}
  \propto
  \vartheta^2
$
\emph{is $\bar{\xi}$-independent} (i.e., insensitive to the atomic density), in accord
with previous reports \cite{Dai_2012,Bruder_2019}.
Moreover, we obtain that for both ${\uvec{k} = {\uvec{x}}, {\uvec{y}}}$,
the ratio
$
  \Re \{S_{\uvec{k}, \parallel}(2\omega_0;2)\}
  /
  \Re \{S_{\uvec{k}, \perp}(2\omega_0;2)\}
$
depends neither on
$\bar \xi$ nor on $\vartheta$ --- a prediction that is yet to be confirmed experimentally.
Both results are due to the fact that all of these signals originate
exclusively in the double-scattering contribution depicted in \fref{fig:scattering} (c), and hence
scale identically with these parameters.

\section{Conclusions}
We developed an original, microscopic, open quantum systems theory of multiple quantum coherence (MQC) signals in fluorescence-detection based measurements on dilute thermal gases. In general, such signals are regarded as sensitive probes of dipole-dipole interactions in many-particle systems \cite{Dai_2012}, but recent experiments \cite{lukas_bruder15} initiated some debate on their true nature \cite{Mukamel_2016,Li_2017,Bruder_2019,kuehn2020}. One of the purposes of this work was to resolve the persisting controversy.

Our theory is characterised by several distinctive features.  First, we 
incorporated the \emph{radiative} part of the dipole-dipole interactions, including the concomitant collective decay processes which are ignored in other treatments ~\cite{Dai_2012,lukas_bruder15,Li_2017,Gao:16,PhysRevLett.120.233401,Yu:19}. Due to the large interatomic distance in dilute atomic vapours, this interaction is here treated perturbatively. 
In this perspective, 2QC signals are sensing the exchange of a single photon within pairs of atoms, and the two-atom correlations resulting from this 
fundamental 
interaction process. Furthermore, we considered a realistic atomic level structure, equipping the atoms with angular momentum, and treated the coupling of such atoms to 
polarized laser pulses of arbitrary strength non-perturbatively. In addition, we effectively accounted for the thermal atomic motion with a configuration average. While it is 
the combination of all these features that allowed us to obtain excellent qualitative, and reasonable quantitative agreement with hitherto partially unexplained experimental 
observations \cite{Bruder_2019,Bruder_phd}, here we would like to highlight the role of \emph{collective decay} processes in the emergence of MQC signals. Such processes 
are completely ignored in treatments which employ the electrostatic form of the 
dipole-dipole interactions~\cite{Dai_2012,lukas_bruder15,Li_2017,Gao:16,PhysRevLett.120.233401,Yu:19}.

In the present contribution, we focused on 1QC and 2QC signals, which we obtained by considering a two-atom model. 
On the one hand, in contrast to previous claims \cite{Mukamel_2016}, we
have shown that the detection of 2QC signals unambiguously indicates the
presence of dipolar interactions---mediated by the exchange of real photons.
On the other hand, our results show that substantial deviations between theory
\cite{Li_2017} and experiment \cite{lukas_bruder15,Bruder_2019} do not
necessitate the consideration of many ($N>2$) interacting particles to predict the
behaviour of 2QC signals; two randomly arranged interacting particles encapsulate
the relevant physics.  

Our method defines a general and versatile framework for the
systematic assessment of higher-order contributions -- which were observed
experimentally \cite{lukas_bruder15,Bruder_2019,Bruder_phd} -- by combining
diagrammatic expansions in the far-field dipole-dipole coupling with
single-atom responses to laser pulses of arbitrary strength
\cite{ketterer2014,binninger19}.
In the future, 
we will explore the nature of the correlations -- quantum or classical -- arising from the exchange of a photon between the atoms.
Finally, we will generalize our method to resolve more subtle features of MQC
spectra \cite{Bruder_phd} due to fine and hyperfine structures of the
involved dipole transitions.

\ack
The authors wish to thank Robert Bennett, Lukas Bruder, Stefan Buhmann, Daniel Finkelstein-Shapiro, Friedemann Landmesser, 
and Frank Stienkemeier for useful and enjoyable discussions. 
E.~G.~C.~acknowledges the support of the G.~H.~Endress Foundation.
V.~S.~and A.~B.~thank the Strategiefonds der Albert-Ludwigs-Universit\"at
Freiburg for partial funding.

\appendix
\section{Exact solution of equation \eref{meq-free}}
\label{app:solve-free-atom}
Given the Liouvillians~\eref{L_a}, we start by rewriting the equation of
motion~\eref{meq-free} as
\begin{equation}
  \dot Q
  =
  \rmi H_{\rm eff}Q
  -\rmi QH^\dagger_{\rm eff}
  +\gamma {\bf D}_\alpha^\dagger\cdot Q{\bf D}_\alpha
  ,
  \label{Ld_Lg}
\end{equation}
where
$
  H_{\rm eff}
  =
  \rmi \gamma/2
  \mathinner{
    {\bf D}_\alpha^\dagger
    \cdot
    {\bf D}_\alpha
  }
$
is a non-Hermitian effective Hamiltonian operator and the operator $\Dd_\alpha$ is given by \eref{dipole_operator}.
Equation \eref{Ld_Lg} admits the following recursive solution (for brevity of
notation, we take the initial time $t_0=0$):
\begin{eqnarray}
  Q(t)
  &=&
  \rme^{\rmi H_{\rm eff}t}
  Q(0)
  \rme^{-\rmi H^\dagger_{\rm eff}t}
  \nonumber\\
  &&+
  \gamma \rme^{\rmi H_{\rm eff}t}
  \Bigl[
    \int_{0}^t \dif{t'}
      \rme^{-\rmi H_{\rm eff}t'}
      {\bf D}^\dagger_\alpha\cdot Q(t'){\bf D}_\alpha
      \rme^{\rmi H^\dagger_{\rm eff}t'}
  \Bigr]
  \rme^{-\rmi H^\dagger_{\rm eff}t}.
  \label{Qt}
\end{eqnarray}
By noticing that ${\bf D}_\alpha^\dagger\cdot {\bf D}_\alpha=\ProjE$ is a
projection operator on the subspace of excited states, we deduce
\begin{equation}
  \rme^{\rmi H_{\rm eff}t}
  =
  \ProjG
  +
  \rme^{- \gamma t/ 2} \ProjE
  =
  \rme^{-\rmi H^\dagger_{\rm eff}t}
  ,
  \label{identities1}
\end{equation}
where $\ProjG=\Id-\ProjE$ is the projector on the ground state.
Employing, in addition, the identities
\begin{equation}
  \ProjG{\bf D}_\alpha^\dagger
  =
  {\bf D}_\alpha^\dagger \ProjE
  =
  \ProjE{\bf D}_\alpha
  =
  {\bf D}_\alpha \ProjG
  =
  0,
  \label{identities2}
\end{equation}
it is easy to establish that only that part of $Q(t')$ in the integrand of
\eref{Qt} which is proportional to $\ProjG$ can yield a contribution.
This component has a trivial time evolution,
$
  \ProjG Q(t') \ProjG
  =
  \ProjG Q(0) \ProjG
$.
Inserting this into \eref{Qt}, with the aid of equations \eref{identities1} and
\eref{identities2} we obtain
\begin{eqnarray}
  Q(t)
  &=&
  \ProjE Q(0) \ProjG
  \mathinner{
    \rme^{- \gamma t /2}
  }
  +
  \ProjG Q(0) \ProjE
  \mathinner{
    \rme^{-\gamma t/2}
  }
  \nonumber \\
  &&+
  \ProjG Q(0) \ProjG
  +
  \ProjE Q(0) \ProjE
  \mathinner{
    \rme^{-\gamma t}
  }
  +
  {\bf D}_\alpha^\dagger\cdot Q(0){\bf D}_\alpha
  \left( 1 - \rme^{-\gamma t} \right)
  ,
  \label{app:Qtsol}
\end{eqnarray}
where, in the last term of the above equation, we have used
${\ProjE{\bf D}_\alpha^\dagger \ProjG = {\bf D}_\alpha^\dagger}$
and
${\ProjG{\bf D}_\alpha \ProjE = {\bf D}_\alpha}$.

\section{Computation of $\bar{\sigma}_{sc}$ from  equation \eref{ell_sc}}
\label{sec:ell_sc}
The detuning-dependent elastic scattering cross-section of a single atom reads \cite{guerin2017}
\begin{equation}
	\sigma_{sc}(\Delta)=\frac{3\lambda_0^2}{2\pi}\frac{1}{1+4\Delta^2/\gamma^2},
	\label{s_sc}
\end{equation}
where $\Delta$ is the detuning between the photon and atomic transition frequencies. For an atom moving with velocity ${\bf v}=(v_x,v_y,v_z)$, the detuning $\Delta={\bf k}\cdot{\bf v}$ is determined solely by the Doppler shift (we assume $\omega_L=\omega_0$). The photon can stem from a laser or be elastically scattered by another atom, i.e., ${\bf k}=(k_x,k_y,k_z)$ and $k=\omega_L/c$. Consequently, $\Delta=\Delta_x+\Delta_y+\Delta_z$ is a random quantity whose components $\Delta_i=k_i v_i$ ($i=x,y,z$) obey the 3-dimensional Maxwell-Boltzmann distribution,
\begin{equation}
	p(\Delta_x,\Delta_y,\Delta_z)=\left(\frac{1}{\bar{\Delta}\sqrt{2\pi}}\right)^3\exp\left(-\frac{\Delta_x^2+\Delta_y^2+\Delta_z^2}{2\bar{\Delta}^2}\right),
\end{equation}
with $\bar{\Delta}$ the root-mean-square Doppler shift. The mean scattering cross-section is then given by the expression
\begin{eqnarray}
	\bar{\sigma}_{sc}&=\int_{-\infty}^{\infty}\dif{\Delta_x}\int_{-\infty}^{\infty}\dif{\Delta_y}\int_{-\infty}^{\infty}\dif{\Delta_z}\sigma_{sc}(\Delta_x,\Delta_y,\Delta_z)p(\Delta_x,\Delta_y,\Delta_z).
	\label{an_sigma}	
\end{eqnarray}
Substituting 
$\lambda_0=\SI{790}{\nano\meter}$, $\gamma=2\pi\times\SI{6}{\mega\hertz}$, and $\bar{\Delta}=\SI{560}{\mega\hertz}$ into \eref{an_sigma}, we obtain
\begin{equation}
	\bar{\sigma}_{sc}\approx 1.14\times \SI{e-16}{m\squared}, 
\end{equation}
which is three orders of magnitude smaller than the resonance scattering cross-section, $\sigma_{sc}(0)=3\lambda_0^2/2\pi\approx 3.0\times \SI{e-13}{m\squared}$.

\section{Complete operator basis and matrix form of equations}
\label{app:basis}
Let us consider a linear master equation of the form
\begin{equation}
	\dot{Q}
	=
	{\cal A} Q
	,
	\label{oper_meq}
\end{equation}
where ${\cal A}$ is a generalized superoperator.
Master equation \eref{meq} is an example of such equations.
Equation \eref{oper_meq} has the formal solution
\begin{equation}
	Q(t)
	=
	\exp({\cal A}t)Q(0)
	.
	\label{sol_oper_meq}
\end{equation}
For individual atoms, such solutions can be obtained in a relatively compact
analytical form (see section \ref{sec:nonint_single}), which simplifies their
physical interpretation.

The superoperator $\exp({\cal A} t)$ in equation \eref{sol_oper_meq} generates 
coupled dynamics of arbitrary atomic operators lumped together in 
a vector ${\bf Q}$.
We choose the latter's elements $Q_n$ to form a complete orthonormal basis, which for a
single atom reads:
\begin{eqnarray}
	{\bf Q}
	&
	=
	&
	(
	\Id/2,\mu_{1}/2,\mu_{2}/2,\mu_{3}/2,
	\sigma_{14},\sigma_{41},
	\sigma_{13},\sigma_{31},
	\sigma_{12},
	\nonumber\\
	&&
	\sigma_{21},
	\sigma_{34},\sigma_{43},
	\sigma_{42},\sigma_{24},
	\sigma_{32},\sigma_{23}
	)^T
	,
	\label{defq}
\end{eqnarray}
where $\sigma_{kl} = \ket{k}\bra{l}$ and
\begin{eqnarray}
	\Id&=&\sigma_{11}+\sigma_{22}+\sigma_{33}+\sigma_{44},\\
	\mu_1&=&\sigma_{22}-\sigma_{33}+\sigma_{44}-\sigma_{11},\\
	\mu_2&=&\sigma_{22}-\sigma_{33}-\sigma_{44}+\sigma_{11},\\
	\mu_3&=&\sigma_{22}+\sigma_{33}-\sigma_{44}-\sigma_{11}.
\end{eqnarray}
The elements of ${\bf Q}$ satisfy the properties of orthonormality,
$\Trace \{Q_m^\dagger Q_n\}=\delta_{nm}$ and completeness,
$O=\sum_{n=1}^{16}c_n Q_n$, with $O$ an arbitrary atomic operator and
expansion coefficients $c_n$.

For $N$ atoms, the complete basis is given by the tensor product
${\bf Q}_1\otimes {\bf Q}_2\otimes\ldots\otimes {\bf Q}_N$.
In this case, vector ${\bf Q}$ has dimension $16^N$, and it evolves according to
\begin{equation}
	{\bf Q}(t)
	=
	{\bf A}(t){\bf Q}(0),
	\label{matr_eq}
\end{equation}
where ${\bf A}(t)=[a_{nm}(t)]$ is a $(16\times 16)^N$ matrix,
whose elements
$a_{nm}(t)=\Trace \{Q_m^\dagger \exp({\cal A}t)Q_n\}$,
with $Q_n$ the elements of the $N$-atom basis set.
Performing a
quantum mechanical average on both sides of equation
\eref{matr_eq}  with respect to the initial atomic density operator, we obtain
\begin{equation}
	\la{\bf Q}(t)\ra
	=
	{\bf A}(t)\la{\bf Q}(0)\ra,
\end{equation}
wherefrom we can express the expectation value of any $N$-atom observable.

\section{Combinations of the tensor T surviving the disorder average}
\label{app:tensorT}
Here we calculate $
\ConfAvg{
	\Tensor{T}_{kl}
	\Tensor{T}{}^{\ast}_{mn}
}
$, that is, we integrate the products $\Tensor{T}_{kl} \Tensor{T}{}^*_{mn}$ over the isotropic angular distribution of the vector $\hat{\bf n}$ connecting two atoms, and replace 
the interatomic distance with its mean value in the far-field limit. 

The matrix elements $\Tensor{T}_{kl}=\hat{\bf e}_k\cdot \Tensor{T}\cdot \hat{\bf e}_l$ are given, according to \eref{tensor_T}, with only its far-field contribution
retained, by
\begin{equation}
  \Tensor{T}_{kl} (\xi, \uvec{n})
  =
 \frac{ 3 \gamma }{4}g(\xi)(\delta_{kl}-\hat{n}_k\hat{n}_l)
  ,
  \label{tensor_T_rewrite}
\end{equation}
with $
  g(\xi)
  =\rmi \rme^{-\rmi \xi}/\xi$ and $\hat{n}_k$ ($k=\{x,y,z\}$) the Cartesian components of the unit vector $\hat{\bf n}=(
  \sin \theta \cos \varphi,
  \sin \theta \sin \varphi,
  \cos \theta
  )$
in spherical coordinates. Hence, 
\begin{equation}
\Tensor{T}_{kl} \Tensor{T}{}^*_{mn}={\rm C}(\xi)(\delta_{kl}\delta_{nm}-\delta_{kl}\hat{n}_n\hat{n}_m-\delta_{nm}\hat{n}_k\hat{n}_l+\hat{n}_k\hat{n}_l\hat{n}_n\hat{n}_m),
\label{tt} 
\end{equation}
where ${\rm C}(\xi)=(3\gamma/4\xi)^2$ is a scalar prefactor. The elementary angular integrals read:
\begin{eqnarray}
 & \int_{0}^\pi &\dif{\theta} \sin \theta \int_0^{2 \pi} \dif{ \varphi}= 4\pi,\label{int_1}\\
  &\int_{0}^\pi &\dif{\theta} \sin \theta \int_0^{2 \pi} \dif{ \varphi}
    \hat{n}_k \hat{n}_l
 =
  \frac{4\pi}{3}
  \delta_{kl}
  ,\label{int_2}\\
&\int_{0}^\pi &\dif{\theta} \sin \theta \int_0^{2 \pi} \dif{ \varphi}  
    \hat{n}_k \hat{n}_l
\hat{n}_m \hat{n}_n
=
\frac{4\pi}{15}
\Bigl(
\delta_{kl} \delta_{mn}
+ \delta_{km} \delta_{ln}
+ \delta_{kn} \delta_{lm}
\Bigr).
\label{int_3}
\end{eqnarray}
Plugging \eref{int_1}-\eref{int_3} into \eref{tt}, and setting $\xi=\bar{\xi}$, we obtain
\begin{equation}
\ConfAvg{
	\Tensor{T}_{kl}
	\Tensor{T}{}^{\ast}_{mn}
}
  =
 {\rm C}(\bar{\xi})\Bigl[\frac{24\pi}{15}  \delta_{kl} \delta_{mn}+
  \frac{4\pi}{15}
  \Bigl(
    \delta_{km} \delta_{ln}
    + \delta_{kn} \delta_{lm}
  \Bigr)\Bigr].
  \label{ang_int}
\end{equation}
Thus, unless the polarization indices $k, l, m, n$ are pairwise identical,
the average
$
  \ConfAvg{
    \Tensor{T}_{kl}
    \Tensor{T}{}^{\ast}_{mn}
  }
$ vanishes. Finally, recalling that $\Tensor{T}=\Tensor{\Gamma}+\rmi\Tensor{\Omega}$, we can easily deduce that the configurational averages 
$
\ConfAvg{
	\Tensor{\Gamma}_{kl}
	\Tensor{\Gamma}_{mn}
}
=
\ConfAvg{
	\Tensor{\Omega}_{kl}
	\Tensor{\Omega}_{mn}
}
=
\ConfAvg{
	\Tensor{T}_{kl}
	\Tensor{T}{}^{\ast}_{mn}
}/2
$, while arbitrary combinations 
$
\ConfAvg{
	\Tensor{\Gamma}_{kl}
	\Tensor{\Omega}_{mn}
}=0
$,
because they include position-dependent phases.

\section*{References}
\bibliography{dip_dip_hooray1}

\end{document}